\documentclass[journal]{IEEEtran}
\usepackage{tabularx,booktabs}
\usepackage{graphicx}
\usepackage{multirow}
\usepackage{amsfonts}
\usepackage{amssymb} 
\usepackage{algorithm}
\usepackage{bm,bbm}
\usepackage{hyperref}
\usepackage{xcolor}
\usepackage[noend]{algpseudocode}
\newcommand\Algphase[1]{%
\vspace*{-0.5\baselineskip}\Statex\hspace*{\dimexpr-\algorithmicindent-2pt\relax}\rule{8.56cm}{0.4pt}
\Statex\hspace*{-\algorithmicindent}\textbf{#1}%
\vspace*{-0.5\baselineskip}\Statex\hspace*{\dimexpr-\algorithmicindent-2pt\relax}\rule{8.56cm}{0.4pt}}

\ifCLASSINFOpdf
\else
\fi

\begin{document}
\title{PSLA: Improving Audio Tagging with Pretraining, Sampling, Labeling, and Aggregation}

\author{Yuan Gong,~\IEEEmembership{Member,~IEEE,}
        Yu-An Chung,~\IEEEmembership{Member,~IEEE,}
        and~James Glass,~\IEEEmembership{Fellow,~IEEE}
\thanks{The authors are with the Computer Science and Artificial Intelligence Laboratory, Massachusetts Institute of Technology, Cambridge, MA 02139 USA (e-mail: yuangong@mit.edu; andyyuan@mit.edu; glass@mit.edu).}}

\markboth{IEEE/ACM TRANSACTIONS ON AUDIO, SPEECH, AND LANGUAGE PROCESSING}%
{Shell \MakeLowercase{\textit{Gong et al.}}: PSLA: Improving Audio Tagging with Pretraining, Sampling, Labeling, and Aggregation}

\maketitle

\begin{abstract}
Audio tagging is an active research area and has a wide range of applications. Since the release of AudioSet, great progress has been made in advancing model performance, which mostly comes from the development of novel model architectures and attention modules. However, we find that appropriate training techniques are equally important for building audio tagging models with AudioSet, but have not received the attention they deserve. To fill the gap, in this work, we present PSLA, a collection of model agnostic training techniques that can noticeably boost the model accuracy including ImageNet pretraining, balanced sampling, data augmentation, label enhancement, model aggregation. While many of these techniques have been previously explored, we conduct a thorough investigation on their design choices and combine them together. By training an EfficientNet with pretraining, balanced sampling, data augmentation, and model aggregation, we obtain a single model (with 13.6M parameters) and an ensemble model that achieve mean average precision (mAP) scores of 0.444 and 0.474 on AudioSet, respectively, outperforming the previous best system of 0.439 with 81M parameters. In addition, our model also achieves a new state-of-the-art mAP of 0.567 on FSD50K. We also investigate the impact of label enhancement on the model performance. Code at \href{https://github.com/YuanGongND/psla}{\color{blue}{https://github.com/YuanGongND/psla}}.
\end{abstract}

\begin{IEEEkeywords}
Audio tagging, Audio event classification, transfer learning, imbalanced learning, noisy label, ensemble
\end{IEEEkeywords}

\IEEEpeerreviewmaketitle

\section{Introduction}

Audio tagging aims to identify sound events that occur in a given audio recording, and enables a variety of Artificial Intelligence-based systems to disambiguate sounds and understand the acoustic environment. Audio tagging has a wide range of health and safety applications in the home, office, industry, transportation, and has become an active research topic in the field of acoustic signal processing. 

In recent years, audio tagging and classification research has moved from small and/or constrained datasets such as ESC-50~\cite{piczak2015esc} and CHiME-Home~\cite{foster2015chime} to much larger datasets with a greater variety and range of real-world audio events and substantially more training data. A  significant milestone in this field occurred with the release of the AudioSet corpus~\cite{gemmeke2017audio} containing over 2 million 10-second audio clips extracted from video and tagged at the utterance level with a set of 527 event labels.  AudioSet is currently the largest and most comprehensive publicly available dataset for audio tagging. Not surprisingly, it has subsequently become the primary source of training and evaluation material for audio tagging research. The availability of AudioSet has encouraged much audio tagging research that has steadily seen the standard evaluation metric of mean average precision (mAP) increase from, for example, 0.314 with shallow fully-connected networks~\cite{gemmeke2017audio}, to 0.392 with a residual network with attention~\cite{ford2019deep} to, most recently, 0.439 with spectrogram and waveform-based convolutional neural networks (CNNs)~\cite{kong2020panns}. In order to cope with the weakly labeled data, multiple instance learning and attention mechanisms have also been the subject of much investigation~\cite{kong2018audio,yu2018multi,kong2019weakly,chen2018class}.

\begin{figure}[t]
  \centering
  \includegraphics[width=8.5cm]{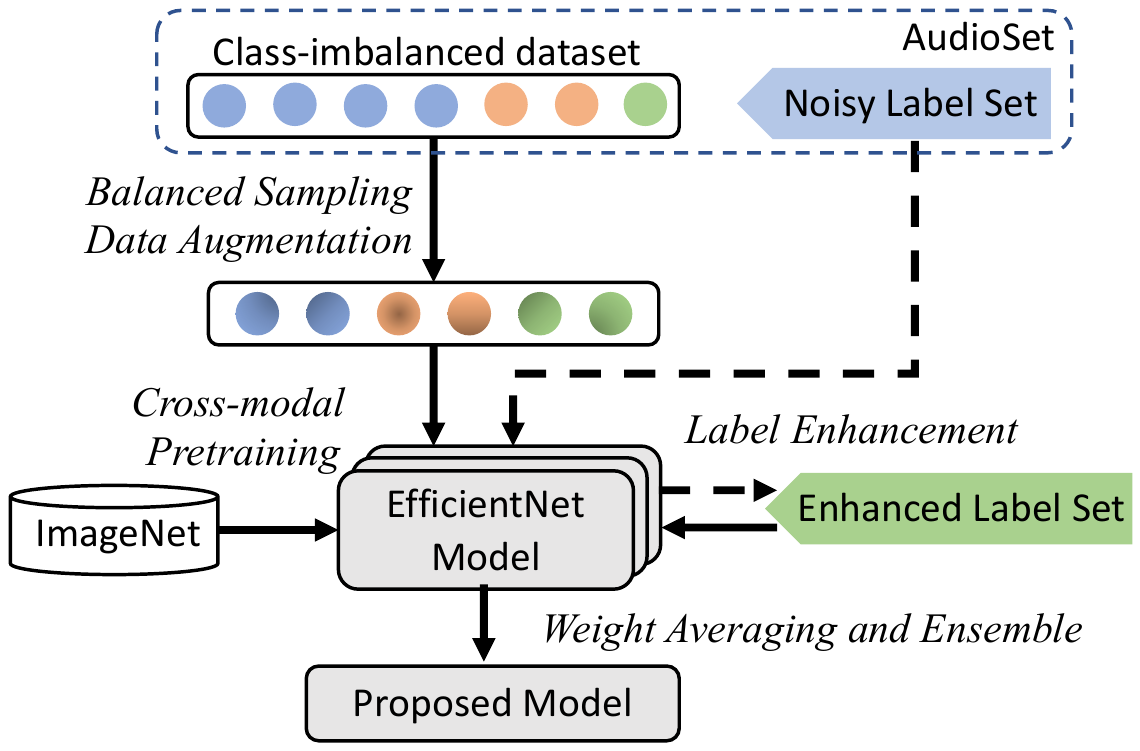}
  \caption{The proposed Pretraining, Sampling, Labeling, and Aggregation (PSLA) training pipeline. AudioSet is extremely class imbalanced and has prevalent annotation errors, we propose a data augmentation/balanced sampling strategy and a label enhancement strategy to alleviate these two problems. We also pretrain the convolutional neural networks with ImageNet and find it leads to a noticeable performance improvement. By further aggregating multiple models with weight averaging and ensemble techniques, we get a model that performs much better than that trained with a conventional pipeline and achieves a new state-of-the-art mAP of 0.474.}
  \label{fig:intro}
\end{figure}

In our audio tagging experiments using Audioset we have observed that, in addition to the particular model architecture being evaluated, significant performance improvements can be achieved via training techniques including cross-modal pre-training, data augmentation, label enhancement, and ensemble modeling. Our empirical evaluations show that these model agnostic techniques lead to significant accuracy improvements, and combining them together can further boost the model accuracy. Specifically, we train an ensemble of EfficientNet~\cite{tan2019efficientnet} models with the proposed set of training techniques and achieve a new state-of-the-art mAP of 0.474 on AudioSet, our single model with 13.6M parameters also achieves an mAP of 0.444, outperforming the previous the best system that contained 81M parameters. In addition, our model also achieves a new state-of-the-art mAP of 0.567 on the FSD50K benchmark~\cite{fonseca2020fsd50k}.


As shown in Figure~\ref{fig:intro}, the training techniques we investigated fall into four main categories. First, we find cross-modal pretraining with ImageNet~\cite{deng2009imagenet} improves the performance of audio tagging CNNs even though AudioSet already contains a substantial amount of in-domain data. Second, we address the Audioset label imbalance by adopting balanced sampling and data augmentation. Third, we observed that there are pervasive annotation errors in AudioSet and studied the impact of such annotation errors on the model performance. We further developed a method to improve training label quality. Fourth, we use weight averaging and ensemble methods to improve the overall performance. Many of these techniques have been proposed previously in isolation.  For example, ImageNet pretraining has been used in~\cite{palanisamy2020rethinking} for small datasets, balanced sampling and data augmentation have been used in~\cite{kong2020panns}, label enhancement has been proposed in~\cite{fonseca2020addressing}, and ensemble modeling has been used in~\cite{ford2019deep,lee2017ensemble,lopez2020ensemble}. To the best of our knowledge however, none of the prior efforts have used more than two of these simultaneously, and the particular implementation is often only briefly mentioned in the literature. In this paper, we thoroughly investigate each of these techniques, a more thorough understanding of the benefits of different training techniques should facilitate a more meaningful comparison between various works because performance differences due to the particular training procedure could overshadow the model architecture or other novel techniques being investigated. The training pipeline we propose is model-agnostic and can serve as a recipe for AudioSet tagging experiments to facilitate fair comparisons with new techniques.

The contributions of this work are summarized as follows:
\begin{enumerate}
    \item We present a collection of training strategies and design choices for audio tagging. We quantify the improvement of each component via extensive experimentation.
    \item By training an ensemble of standard EfficientNet models with the proposed training procedure, we achieve a new state-of-the-art mAP of 0.474 on AudioSet, outperforming the best previous system of 0.439.
    \item We release the code, model, and enhanced label set. The training pipeline can serve as a recipe of AudioSet training to facilitate future audio tagging research.
\end{enumerate}

The paper is organized as follows. We first describe the baseline model architecture in Section~\ref{sec:baseline}, then we gradually improve the baseline model performance on AudioSet by adding new training techniques in Sections~\ref{sec:pretrain},~\ref{sec:dataaug},~\ref{sec:label}, and~\ref{sec:ensemble}. In each section, we first review the corresponding technique and then present our implementation and results. We present an ablation study, experiments on FSD50K and other model architectures, and a discussion of the results in Section~\ref{sec:ablation}. We conclude the paper in Section~\ref{sec:discuss}.

\section{Experiment Setting and Baseline Model}
\label{sec:baseline}
\subsection{Dataset}

\begin{table}[]
\centering
\caption{The AudioSet~\cite{gemmeke2017audio} Statistics.}
\label{tab:audioset}
\begin{tabular}{@{}lccc@{}}
\toprule
                        & Balanced Train & Full Train & Evaluation \\ \midrule
AudioSet                & 22,176         & 2,065,161             & 20,383     \\
Downloaded & 20,785         & 1,953,082               & 19,185     \\
Downloaded Ratio        & 93.7\%         & 94.6\%                  & 94.1\%     \\ \bottomrule
\end{tabular}
\end{table}

In this work, we mainly focus on AudioSet~\cite{gemmeke2017audio}, a collection of over 2 million 10-second audio clips excised from YouTube videos and labeled with the sounds that the clip contains from a set of 527 labels. 
AudioSet is a weakly labeled and multi-label dataset, i.e., labels are given to a clip with no indication of where in the clip the associated sound occurred, and every clip can, and often does, have multiple labels associated with it. As shown in Table~\ref{tab:audioset}, the dataset is split into three subsets: balanced train, unbalanced train, and evaluation. In this paper, we combine the balanced and unbalanced training set as the full training set. The balanced train dataset is a set of 22,176 recordings, where each class has at least 49 samples, while the full train set contains the entire 2 million recordings.
The evaluation set consists of 20,383 recordings and contains at least 59 examples for each class.  To obtain the raw audio, we extracted the dataset from YouTube.  Due to the constant change in video availability (e.g., videos being removed, taken down) there is a natural shrinkage (about 5\%) from the original dataset~\cite{gemmeke2017audio}. Specifically, we downloaded 20,785 (94\%), 1,953,082 (95\%), and 19,185 (94\%) recordings for the balanced train, full train, and evaluation set, respectively, which is consistent with previous literature (e.g., \cite{kong2020panns}). Therefore, we do make fair comparisons with previous state-of-the-art models by evaluating on the same subset of the evaluation dataset.

We also evaluate the proposed PSLA training framework on FSD50K~\cite{fonseca2020fsd50k}, a recently collected data set of sound event audio clips with 200 classes drawn from the AudioSet ontology to see how the PSLA framework generalizes. FSD50K contains 37,134 audio clips for training, 4,170 audio clips for validation, and 10,231 audio clips for evaluation. The audio clips are of variable length from 0.3 to 30s with an average of 7.6s (7.1s for the training and validation set, 9.8s for the evaluation set). For both AudioSet and FSD50K, we sample the audio at 16kHz.

\begin{figure}
  \centering
  \includegraphics[width=8.5cm]{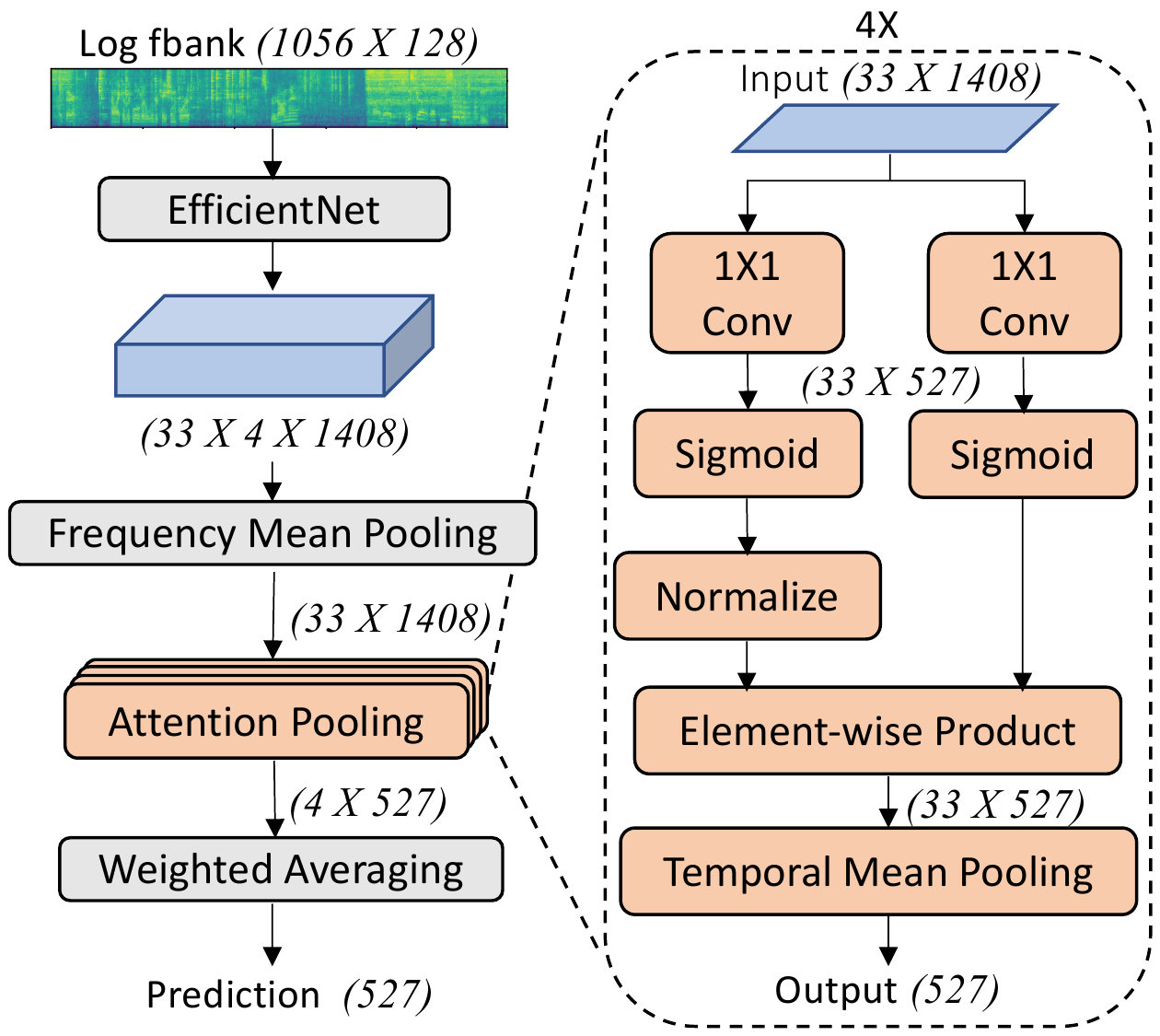}
  \caption{The audio tagging model used in this work. The 10-second waveform is first converted to a $1056\times128$ log Mel filterbank (fbank) feature vector and input to the EfficientNet model. The output of the penultimate layer of EfficientNet is a $33\times4\times1408$ tensor. We apply a frequency mean pooling to produce a $33\times1408$ representation that is fed into a 4-headed attention pooling module. In each head, the CNN output is transformed into a $33\times527$ dimensional tensor via a set of 1$\times$1 convolution layers with a parallel attention branch and classification branch. We multiply the output of each branch element-wise and apply a temporal mean pooling (implemented by summation). Finally, we sum the weighted output of each attention head after it has been scaled by a learnable weight and produce the final prediction for all classes.}
  \label{fig:model}
\end{figure}

\subsection{Training and Evaluation Details}
\label{sec:train}
For all AudioSet experiments in this paper, we train the neural network model with a batch size of 100, the Adam optimizer~\cite{kingma2015adam}, and use binary cross-entropy (BCE) loss. We use a fixed initial learning rate of 1e-3 and 1e-4 and cut it in half every 5 epochs after the $35^{th}$ and $10^{th}$ epoch for all balanced set and full set experiments, respectively. The reason why a smaller learning rate is used for the full AudioSet is that the full set is about 100 times larger than the balanced set, using a smaller learning rate can avoid the model falling into a local minima before it sees all samples. We use a linear learning rate warm-up strategy for the first 1,000 iterations. As in previous efforts, we train the model with 60 and 30 epochs for all balanced set and full set experiments, respectively, and report the mean result on the evaluation set of the last 5 epochs.

We use the mean average precision (mAP) of all the classes as our main evaluation metric since it is the most commonly used audio tagging evaluation metric on AudioSet. Mean average precision is an approximation of the area under a class’s precision-recall curve, which is more informative of performance when dealing with imbalanced datasets such as AudioSet and FSD50k compared with the average area under the curve of the receiver operating characteristic curve~\cite{bradley1997use,davis2006relationship}. In the discussion section, we also report the average area under the curve (AUC) of the receiver operating characteristic curve and sensitivity index (d-prime) in order to compare our model with previous work that only reports AUC and d-prime.

\subsection{Baseline Model}
\label{sec:base_mdl}

In this work, we use a similar model structure as in~\cite{ford2019deep}, illustrated in Figure~\ref{fig:model}.  Each 10-second audio waveform is first converted to a sequence of 128 dimensional log Mel filterbank (fbank) features computed with a 25ms Hamming window every 10ms. We conduct zero padding to make all audio clips have 1056 frames. This results in a $1056\times128$ feature vector that is input to a CNN model. In~\cite{ford2019deep} the CNN was based on the ResNet50 model~\cite{he2016deep}.  In our work, the CNN is based on the EfficientNet-B2 model~\cite{tan2019efficientnet} since it requires a smaller number of parameters and is faster for training and inference.  The EfficientNet model effectively downsamples the time and frequency dimensions by a factor of 32.  The  penultimate output of the model is a $33\times4\times1408$ tensor.  We apply mean pooling over the 4 frequency dimensions to produce a $33\times1408$ representation that is fed into a multi-head attention module.  The attention module consists of an attention branch and a classification branch.  Each branch transforms the CNN mean pooled output into a $33\times527$ dimensional tensor via a set of $1\times1$ convolutional filters.  After a sigmoid non-linearity and a normalization on the attention branch, we combine the two branches via a element-wise product.  A temporal mean pooling (implemented by summation) is then performed to produce a final $527$ dimensional output for each class label.  Unlike~\cite{ford2019deep}, we use a 4-headed attention module instead of a single-head one in this work.  We sum the weighted output of each attention head after it has been scaled by a learnable weight to produce the final output.


EfficientNet~\cite{tan2019efficientnet} is a recent proposed convolutional neural network architecture that has shown an advantage on both accuracy and efficiency over previous architectures. Such advantage mainly comes from two design: First, EfficientNet is based on the mobile inverted bottleneck convolution (MBConv) block~\cite{sandler2018mobilenetv2,tan2019mnasnet}, an efficient residual convolution block. Second, EfficientNet scales the network on all dimensions (i.e., width, depth, and input resolution), which is demonstrated to be a better strategy than scaling only one dimension. In this work, we use EfficientNet-B2 that consists of 9 stages, 339 layers. The original EfficientNet-B2 model for image classification has 9.11M parameters, after adding the attention module and adjusting the classification layer, our audio tagging model has 13.64M parameters in total. As shown in Table~\ref{tab:baseline}, the EfficientNet model achieves slightly worse performance than the ResNet-50 model, but has 12 million fewer parameters. In the rest of the paper, we keep using the EfficientNet model and show that a significant improvement can be achieved without modifying its model architecture.

\begin{table}[]
\centering
\caption{Mean Average Precision (mAP) Comparison of the ResNet Model~\cite{ford2019deep} and the EfficientNet Model Used in This Paper.}
\label{tab:baseline}
\begin{tabular}{@{}cccc@{}}
\toprule
                & \# Parameters & Balanced Set & Full Set \\ \midrule
ResNet-50       &  25.66M            &    0.1635     &     0.3790   \\
EfficientNet-B2 &  13.64M             &    0.1570     &  0.3723      \\ \bottomrule
\end{tabular}
\end{table}

\section{Network Pretraining}
\label{sec:pretrain}

Transfer learning and network pretraining have been widely used in computer vision, natural language processing, speech and audio processing in recent years~\cite{devlin2019bert,he2019rethinking,chung2019unsupervised}. The typical process is to first train a model with either a large out-of-domain or unlabeled dataset using an auxiliary task and then fine-tune the model with in-domain data for the main task. The idea being that the knowledge learned from the pretraining task can be transferred to the main task. 

For the audio tagging task, both supervised pretraining (e.g., in~\cite{kong2020panns}) and self-supervised pretraining (e.g., in~\cite{saeed2020contrastive,shor2020towards,tagliasacchi2020pre,jansen2018unsupervised,wang2020contrastive}) using audio data have been studied in recent years.  Performance improvement is typically achieved when the in-domain dataset is small (e.g., ESC-50~\cite{piczak2015esc}, UrbanSound~\cite{salamon2014dataset}, and balanced AudioSet). However, it has not been reported that a pretrained model can outperform a state-of-the-art audio tagging model trained from scratch using the full AudioSet, possibly because the full AudioSet contains 2 million audio recordings and there is no larger annotated dataset available. While theoretically self-supervised pretraining can leverage an unlimited amount of unlabelled audio data, in practice it takes effort to find and process large scale data with sufficient variety and coverage of the 527 sound classes.

\begin{table}[t]
\centering
\caption{Performance Impact on mAP Due to \\ Pretraining with ImageNet Data.}
\label{tab:pretrain}
\begin{tabular}{@{}ccc@{}}
\toprule
                             & Balanced Set & Full Set \\ \midrule
No pretraining &  0.1570                  &           0.3723         \\
With pretraining &    0.2385                  &           0.3939         \\ \bottomrule
\end{tabular}
\end{table}

\begin{figure}[t]
  \centering
  \includegraphics[width=8.5cm]{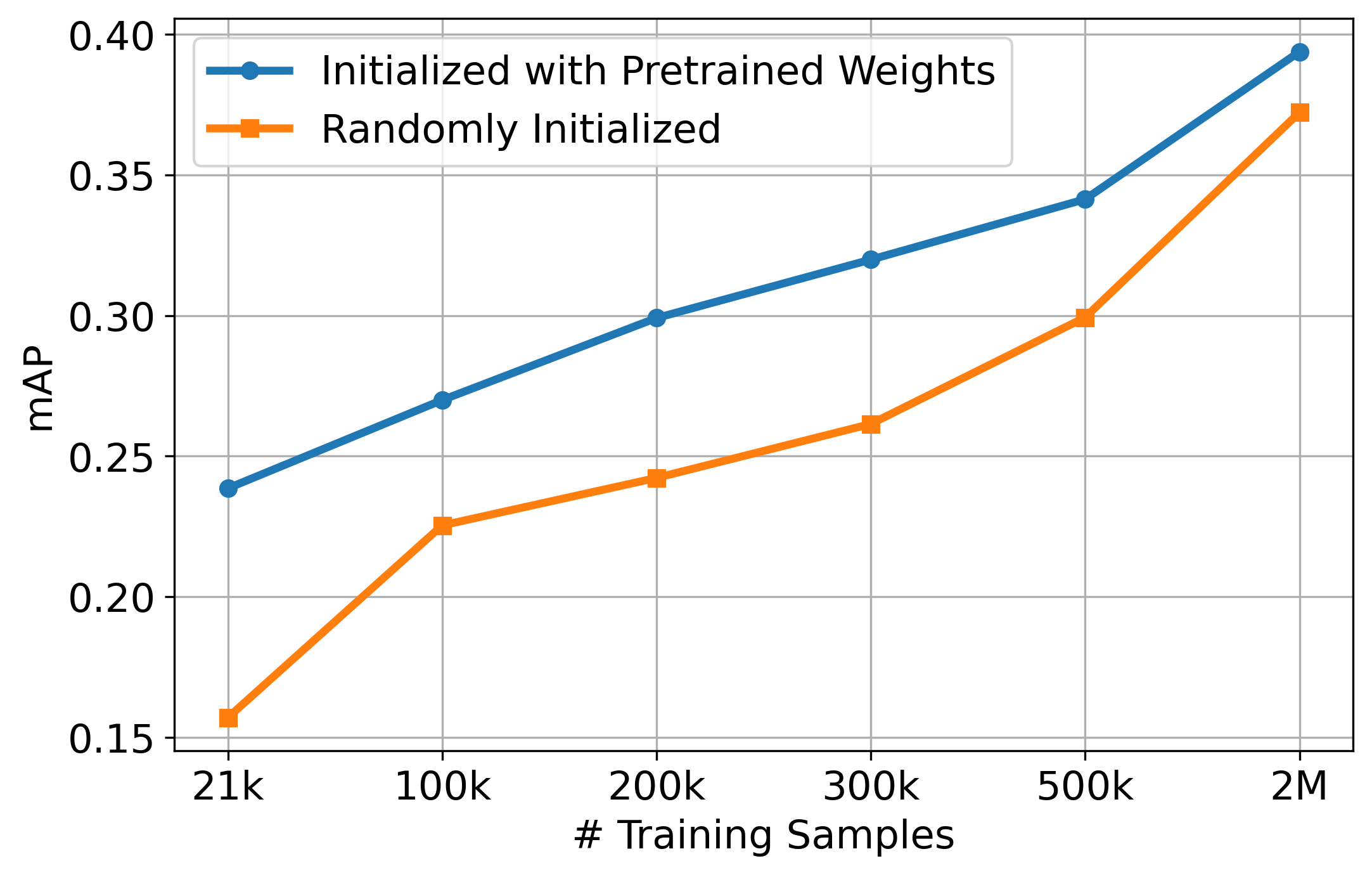}
  \caption{Comparison of the performance of ImageNet-pretrained model and random-initialized model with different training data volume.}
  \label{fig:pretrain}
\end{figure}

In contrast to the above-mentioned efforts, we find noticeable performance improvement can be achieved by pretraining the CNN with the ImageNet dataset~\cite{deng2009imagenet} used for visual object classification, even when the training data for the end task of audio tagging is the full AudioSet. In our experiment, we initialize the EfficientNet (the second to the penultimate layer) with 1) ImageNet-pretrained weights (released by the authors of~\cite{tan2019efficientnet}), and 2) random weights (He Uniform initialization~\cite{he2015delving}). We then train both models in exactly the same way as described in Section~\ref{sec:train}. 

As shown in Table~\ref{tab:pretrain}, ImageNet pretraining leads to a 51.9\% and 5.8\% relative improvement for the balanced set and full set experiment, respectively. To see the relationship between the performance improvement and the end-task training data volume, we further evaluate the performance when the audio tagging training data volume is 100k, 200k, 300k, and 500k (all comprised of the entire balanced set and samples randomly taken from the full set). As shown in Figure~\ref{fig:pretrain}, the performance improvement decreases with the training data volume, but is always noticeable. In addition, we find the performance improvement led by ImageNet pretraining is much larger than that led by more training iterations, e.g., when trained with the balanced AudioSet, the model trained with 120 epochs achieves an mAP of 0.1694, which is only slightly better than the model trained with 60 epochs and is significantly worse than the model trained with ImageNet pretraining that achieves an mAP of 0.2385.

In some sense, it is surprising that pretraining a model with data from a different modality can be effective.  However, transfer learning from computer vision tasks to audio tasks is not new and has been previously studied in~\cite{gwardys2014deep,guzhov2020esresnet,adapa2019urban,palanisamy2020rethinking}.  However, we believe this is the first time it has been demonstrated to be effective when the dataset of the audio task is at this scale, indicating the auxiliary image classification task helps the model learn some complementary knowledge. We hypothesize that the improvements may be due to the model learning to recognize low-level features such as edges during pretraining.  Such knowledge could potentially be relevant for finding acoustic ``edges" in the spectrogram. 

In practice, many commonly used CNN architectures (e.g., Inception~\cite{szegedy2015going}, ResNet~\cite{he2016deep}, EfficientNet~\cite{tan2019efficientnet}) have off-the-shelf ImageNet-pretrained models for both TensorFlow and PyTorch. It is also straightforward to adapt these off-the-shelf models to audio tasks. The only things that need to be modified are the first convolution layer and the last classification layer. Since the input of vision tasks is a 3-channel image while the input to the audio task is a single-channel spectrogram, we adjust the input channel of the first convolutional layer from 3 to 1 and initialize it with random weights. Since the classification task is essentially different, we abandon the last classification layer of the pretrained model and feed the output of the penultimate layer to our succeeding layers. We implement this using the  \texttt{efficientnet\_pytorch}\footnote{https://github.com/lukemelas/EfficientNet-PyTorch} package.

In summary, the advantages of using ImageNet pretraining are as follows. First, no additional in-domain labeled or unlabeled datasets are needed. This is important because currently there is no audio tagging dataset of comparable size to AudioSet. Second, ImageNet pretraining can lead to consistent performance improvement even when the in-domain training data size is huge. Third, ImageNet pretraining is practically easy to implement. The limitation is that it is only applicable to models that take 2D image-like input (e.g., spectrogram). Nevertheless, a majority of deep learning models for audio tasks do fall in this category. In the following sections, we use Imagenet pretraining by default for all experiments.

\section{Balanced Sampling and Data Augmentation}
\label{sec:dataaug}

\subsection{Balanced Sampling}
\label{sec:bal}

\begin{figure}
  \centering
  \includegraphics[width=8.5cm]{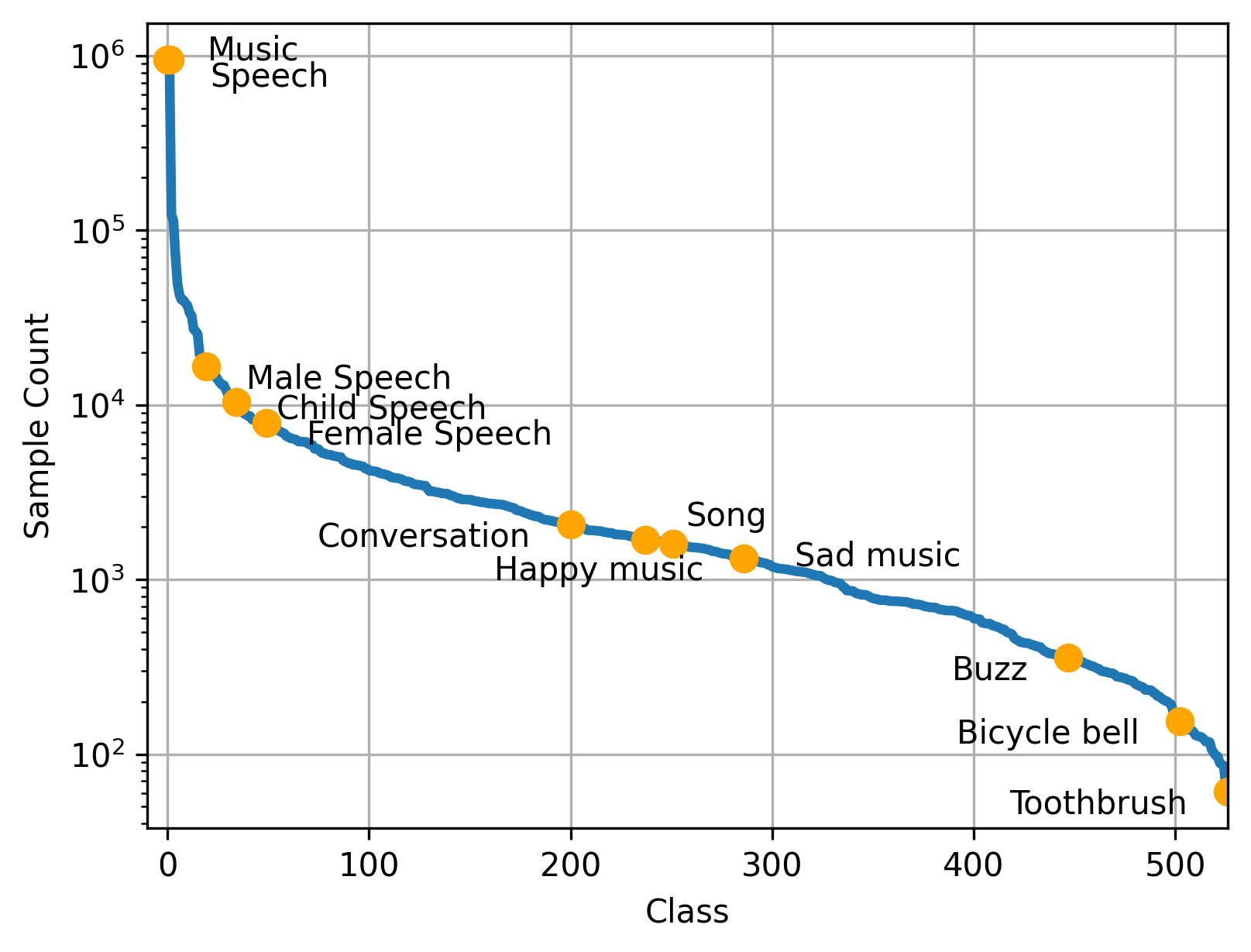}
  \caption{Sample count of each class in the full AudioSet (vertical axis is in log scale). Note that the sample count of the ``Speech'' class is substantially larger than the sum of sample counts of the ``Male Speech'', ``Female Speech'', and ``Child Speech'' class. Similarly, the sample count of the ``Music'' class is substantially larger than the sum of sample counts of the ``Happy music'' and ``Sad music'' class. This indicates a potential prevalent miss annotation issue in AudioSet.}
  \label{fig:samplecnt}
\end{figure}

As might be expected, the frequency of occurrence of different sound events ranges widely. It is not surprising then that a large scale audio tagging dataset is class imbalanced. As shown in Figure~\ref{fig:samplecnt}, the most frequent AudioSet class is ``Music'' which has 949,029 samples, while the most infrequent class ``Toothbrush'' only has 61 samples, leading to a ratio of 15,557. Such imbalances can have a large impact on performance, particularly for low-frequency classes~\cite{he2009learning}. 

\begin{table}[t]
\centering
\caption{Performance Impact on mAP Due to Various Balanced Sampling and Data Augmentation Strategies.}
\label{tab:aug}
\begin{tabular}{@{}lcc@{}}
\toprule
                                      & Balanced Set & Full Set \\ \midrule
Baseline &   0.2385 &          0.3939             \\
+ Balanced Sampling  &   -                   &   0.3721               \\
+ Time-Frequency Masking  &  0.2818                    &        0.4265                \\
+ Mix-up Training &   0.3108                   &       0.4397                \\ \bottomrule
\end{tabular}
\end{table}

\begin{figure}[h]
  \centering
  \includegraphics[width=7cm]{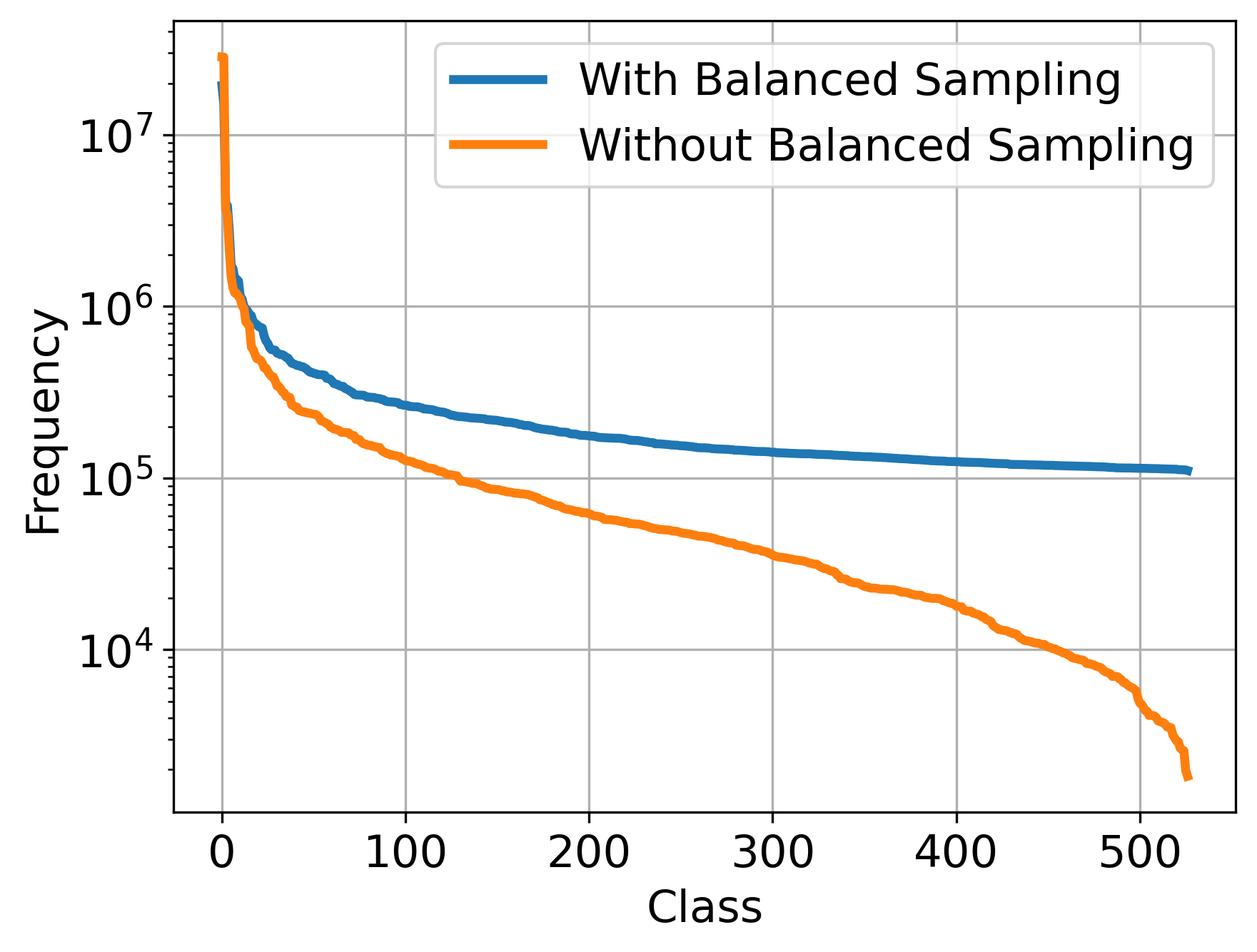}
  \caption{Sorted sampled frequency of each class after 30 training epochs.}
  \label{fig:class_freq}
\end{figure}

With such large data imbalance, simple upsampling or downsampling are difficult to implement because upsampling will make the dataset unacceptably large while downsampling will waste a large portion of the data. Moreover, AudioSet is a multi-label dataset, making it even harder to implement up/downsampling methods. In this work, we propose a random balanced sampling method to alleviate the class imbalance problem. Note that balanced sampling on AudioSet has been used in~\cite{kong2018audio,kong2019weakly,kong2020panns}, but is only briefly mentioned and the details can only be found in the source code.

\begin{algorithm}[!t]
\caption{Balanced Sampling and Data Augmentation}
\label{alg:1}
\begin{algorithmic}[1]
\Require{
\quad \newline
Multi-label Dataset $\mathcal{D}=\{\boldsymbol{x}^{(i)}, \boldsymbol{y}^{(i)}\}$, $i \in \{1,...,N\}$}
\Algphase{Procedure 1: Generate Sampling Weight}
\textbf{Input:}  Label Set $\{\boldsymbol{y}^{(i)}\}$ \newline
\textbf{Output:}  Sample Weight Set $\mathcal{W}=\{\boldsymbol{w}^{(i)}$\}, $i \in \{1,...,N\}$
\State traverse $\{\boldsymbol{y}^{(i)}\}$, count sample number $\boldsymbol{c_k}$ of each class $k$
\State initialize $\boldsymbol{w}^{(i)}=0$, $i \in \{1,...,N\}$
\For {each sample $i$}
\For {each class $k \in \boldsymbol{y}^{(i)}$}
\State $\boldsymbol{w}^{(i)} = \boldsymbol{w}^{(i)} + 1/c_k$
\EndFor 
\EndFor
\Return $\mathcal{W}=\{\boldsymbol{w}^{(i)}\}$

\Algphase{Procedure 2: Sampling and Augmentation in Training}
\textbf{Input:} $\{\boldsymbol{x}^{(i)}, \boldsymbol{y}^{(i)}\}$, $\mathcal{W}$, $F$, $T$, $M$
\For{every epoch}
\For{$n \in \{1,...,N\}$ }
\State draw $i\sim multinomial (\mathcal{W})$
\If{$unif(0,1)<$ mix\-up rate $M$}
\State draw $j\sim unif \{1,N\}$
\State draw $\lambda\sim Beta(\alpha,\alpha)$
\State $x=\lambda x^{(i)}+(1-\lambda)x^{(j)}$
\State $y=\lambda y^{(i)}+(1-\lambda)y^{(j)}$
\Else
\State $x=x^{(i)}$, $y=y^{(i)}$
\EndIf
\State draw $f\sim unif(0, F)$, $f_0\sim unif(0, 128-f)$
\State draw $t\sim unif(0, T)$, $t_0\sim unif(0, 1056-t)$
\State $x=Masking(f_0, t_0, f, t)(x)$
\State use $(x,y)$ to train the neural network
\EndFor
\EndFor
\end{algorithmic}
\end{algorithm}

The proposed random balanced sampling approach is shown in Algorithm~\ref{alg:1}, lines 1-8. We first count the sample number $c_k$ of each class $k$ over the entire dataset. We then assign a sampling weight for each sample, specifically, the weight $\boldsymbol{w}^{(i)}$ of the $i^{th}$ sample is $\sum_{k=1}^{527} \mathbbm{1}_{\{k\in \boldsymbol{y}^{(i)}\}} 1/c_k$. This assigns a higher weight for samples containing rare audio events and also takes all audio events that appear in the sample into consideration. During training, we still feed $N$ samples ($N$ is the dataset size) to the model for each epoch, but instead of traversing the dataset, we draw a sample from the multinomial distribution parameterized by the above-mentioned sample weights with replacement. That makes rare sound event samples more likely to be seen by the model. The advantages of the proposed random sampling are 1) it is a compromise of upsampling and downsampling. It wastes fewer samples than downsampling while keeping the number of $N$ samples fed to the model every epoch; 2) it is applicable to multi-label datasets; and 3) the model sees a different set of data every epoch, so the model checkpoints after every epoch have a greater diversity, which is helpful for ensembles~\cite{efron1994introduction,breiman1996bagging}, as we will discuss in  Section~\ref{sec:ensemble}.  

\begin{figure}
  \centering
  \includegraphics[width=8.5cm]{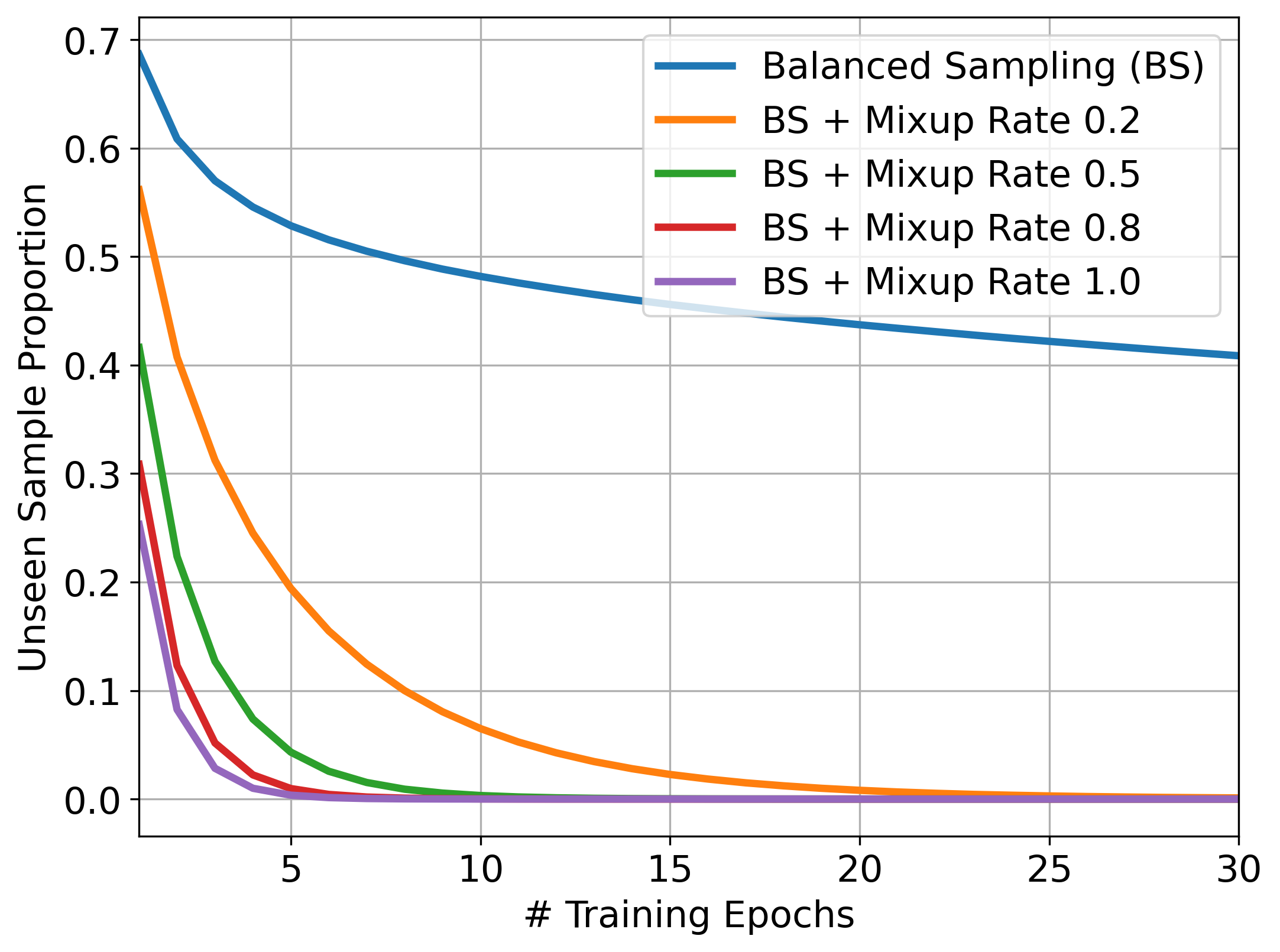}
  \caption{The proportion of unseen samples with the training epochs. Mixup rate is the probability that the sample input to the model is a mixed-up sample. In our implementation, one of the two mixed-up samples is drawn from a uniform distribution, while the other is drawn using the balanced sampling multinomial distribution.}
  \label{fig:waste}
\end{figure}

As shown in Figure~\ref{fig:class_freq}, while the proposed balanced sampling algorithm greatly alleviates the data imbalance issue, the sampled frequency of each class is still imbalanced after the balanced sampling algorithm is applied. This is because AudioSet is a multi-label dataset and minority classes are usually paired with majority classes, thus oversampling the minority class also directly oversamples the majority class. We compare the performance of the model trained with plain dataset traversal (with data reshuffled at every epoch) and with the proposed random sampling. As shown in Table~\ref{tab:aug}, we find random balanced sampling actually lowers the performance. This result is not surprising because: 1) while better than downsampling, there is still a substantial amount of data wasted every epoch. As shown in Figure~\ref{fig:waste}, 40.9\% data is not seen by the model after 30 training epochs; 2) while the low-frequency class samples and high-frequency class samples are roughly equally seen by the model, the low-frequency class samples are actually repeated samples. Both issues increase the risk of model overfitting. Therefore, we explored the use of data augmentation to overcome this problem.

\subsection{Time and Frequency Masking}

We first consider simple time and frequency masking for data augmentation, which has been found to be effective for audio tagging~\cite{kong2020panns} and speech recognition~\cite{park2019specaugment}. Frequency masking is applied so that $f$ consecutive frequency channels [$f_0$, $f_0$ + $f$) are masked, where $f\sim unif(0, F)$, $f_0\sim unif(0, 128-f)$, and $F$ is the maximum possible length of the frequency mask. Similarly, time masking is applied so that $t$ consecutive frequency channels [$t_0$, $t_0$ + $t$) are masked, where $t\sim unif(0, T)$, $t_0\sim unif(0, 1056-t)$, and $T$ is the maximum possible length of the frequency mask. Note that 128 and 1056 are the input dimensions of our model. We use the implementation of \texttt{torchaudio.transforms.FrequencyMasking} and \texttt{TimeMasking}, $F=48$ and $T=192$. The masking parameters $f_0,t_0,f,t$ are sampled on-the-fly for each audio sample during training to minimize the chance of repeated audio samples being fed to the model. As shown in Table~\ref{tab:aug}, time and frequency masking improves audio tagging performance considerably, with relative improvements of 18.2\% and 14.6\% achieved for the balanced set and full set experiment, respectively. Note that the overall amount of training samples per epoch remains the same.  We hypothesize that the effectiveness of masking is due to the reduction of repeated samples in the training data, especially for low-frequency samples.

\subsection{Mix-up Training}

An additional form of data augmentation we explored is called \emph{mix-up training} where weighted combinations of audio samples are combined to make new samples. Mix-up training creates convex combinations of pairs of examples and their corresponding labels. Studies have shown it can improve the performance of image classification, voice command recognition~\cite{tokozume2018between,zhang2018mixup}, and audio tagging~\cite{kong2020panns,tokozume2018learning}. Specifically, mix-up training constructs augmented training examples as follows:

\[x=\lambda x^{(i)}+(1-\lambda)x^{(j)}\]
\[y=\lambda y^{(i)}+(1-\lambda)y^{(j)}\]

\begin{table}[t]
\centering
\caption{Performance as a function of Mix-up Rate\\ (Training on Balanced Set with $\alpha=10$).}
\label{tab:mixuprate}
\begin{tabular}{@{}lccccc@{}}
\toprule
Mixup Rate & 0      & 0.2    & 0.5    & 0.8    & 1.0    \\ \midrule
mAP        & 0.2818 & 0.3060 & 0.3108 & 0.3119 & 0.2928 \\ \bottomrule
\end{tabular}
\end{table}

\begin{table}[h]
\centering
\caption{Performance as a function of $\alpha$ \\(Training on balanced set with Mix-up rate$=0.5$).}
\label{tab:beta}
\begin{tabular}{@{}lcccc@{}}
\toprule
$\alpha$ & $-\infty$ & 0.1    & 1      & 10     \\ \midrule
mAP      & 0.2818    & 0.3004 & 0.3087 & 0.3108 \\ \bottomrule
\end{tabular}
\end{table}

where $x^{(i)}$ and $x^{(j)}$ are two different training audio samples, $y^{(i)}$ and $y^{(j)}$ are the corresponding labels, $\lambda\in[0,1]$ and $x$ is the mixed-up new audio sample, and $y$ is the resulting label. We conduct mix-up on the waveform level.

Past explanations for why mix-up training improves performance include:  1) it increases the variation of the training data~\cite{kong2020panns,tokozume2018learning}; 2) it leads to an enlargement of Fisher’s criterion in the feature space and a regularization of the positional relationship among the feature distributions of the classes~\cite{tokozume2018between,tokozume2018learning}; and 3) it reduces the model's memorization of corrupt labels~\cite{zhang2018mixup}.

In addition to these observations, we find mix-up training has an additional advantage for imbalanced datasets. As we discussed in Section~\ref{sec:bal}, balanced sampling, while making the low-frequency class samples more prevalent, has the unfortunate side effect of wasting a large number of (40.9\%) class samples. By adopting the mixup strategy, the model can see twice the number of samples within the same training epoch. This advantage can be increased if one of the two mixed-up samples is drawn from a uniform distribution, while the other is drawn using the balanced sampling multinomial distribution introduced in the previous section. Intuitively, mixing up a rare sound event (e.g., toothbrush) with a frequent one (e.g., music) is more reasonable than mixing up two rare sound events. Some previous synthetic audio event detection datasets use a similar method to construct samples~\cite{mesaros2017dcase}. As shown in Figure~\ref{fig:waste}, the mix-up strategy can reduce the unseen samples to almost zero in just a few epochs.

We further make two modifications based on previous efforts. In prior work $\lambda$ is drawn from a uniform distribution $unif(0,1)$~\cite{tokozume2018learning} or Beta distribution $Beta(\alpha,\alpha)$ with $\alpha<1$~\cite{zhang2018mixup}, where 
\[Beta(\alpha,\alpha):\,\, prob(x;\alpha,\alpha)=\frac{x^{\alpha-1}(1-x)^{\alpha-1}}{B(\alpha,\alpha)}\label{eq:BetaDensity}\]
where $B$ is the beta function
\[B(\alpha,\alpha)=\int_{0}^{1}t^{\alpha-1}(1-t)^{\alpha-1}dt\]

Thus $\lambda$ has a relatively high likelihood to be close to either 0 or 1. From the perspective of sound mixing and reducing the number of unseen samples, a $\lambda$ close to 0.5 could be more reasonable since it leads to more ``evenly'' mixed up samples and the model can  see both samples. 
Second, since samples in the evaluation set are not mixed up, mixing up all samples during training might lead to a gap between training and evaluation. Thus we set a mix-up {\it rate} to control the number of samples to mix up during training, a mixup rate of 0.5 means that 50\% training samples are mixup samples and the rest 50\% training samples are non-synthetic samples. Therefore, the model can see non-synthetic samples during training. As shown in Figure~\ref{fig:waste}, a mix-up rate of 0.5 results in 95\% samples being seen by the model in 5 epochs. For non mix-up samples, the data loader only needs to load one audio sample instead of two. A low mix-up rate can also reduce the data loading and pre-processing cost during training, which is non-negligible because it is almost impossible to fit the full AudioSet into memory. 

We evaluate the impact of mix-up rate and $\alpha$, as shown in Tables~\ref{tab:mixuprate} and~\ref{tab:beta}. A larger $\alpha$ and a medium mix-up rate indeed lead to better classification performance. Combining them achieves 0.3108 mAP, which is better than a plain setting of $\alpha$=mixup rate=$1$ that achieves 0.3079 mAP. We use $\alpha=10$ and mix-up rate$=0.5$ in all subsequent experiments.

\subsection{Summary}

We combine the balanced sampling and masking and mix-up data augmentation strategies together, as described in Algorithm~\ref{alg:1}. We summarize the contribution of each component in Table~\ref{tab:aug}. It is worth mentioning that while balanced sampling alone lowers the performance, it is helpful when combined with data augmentation strategies. By adopting balanced sampling and data augmentation, an 11.6\% relative improvement and an mAP of 0.4397 are achieved for the full set experiment. We only do data augmentation for balanced set experiments as the data is already roughly balanced and obtain a 30.3\% relative improvement and an mAP of 0.3108, demonstrating the effectiveness of data augmentation for small datasets. Finally, it is worth mentioning that by merely adopting ImageNet pretraining, balanced sampling, and data augmentation with a standard EfficientNet architecture, the model already outperforms the previous best system. In the following sections, we use balanced sampling (for the full AudioSet) and data augmentation as defaults for all experiments.

\begin{figure}[t]
  \centering
  \includegraphics[width=8.5cm]{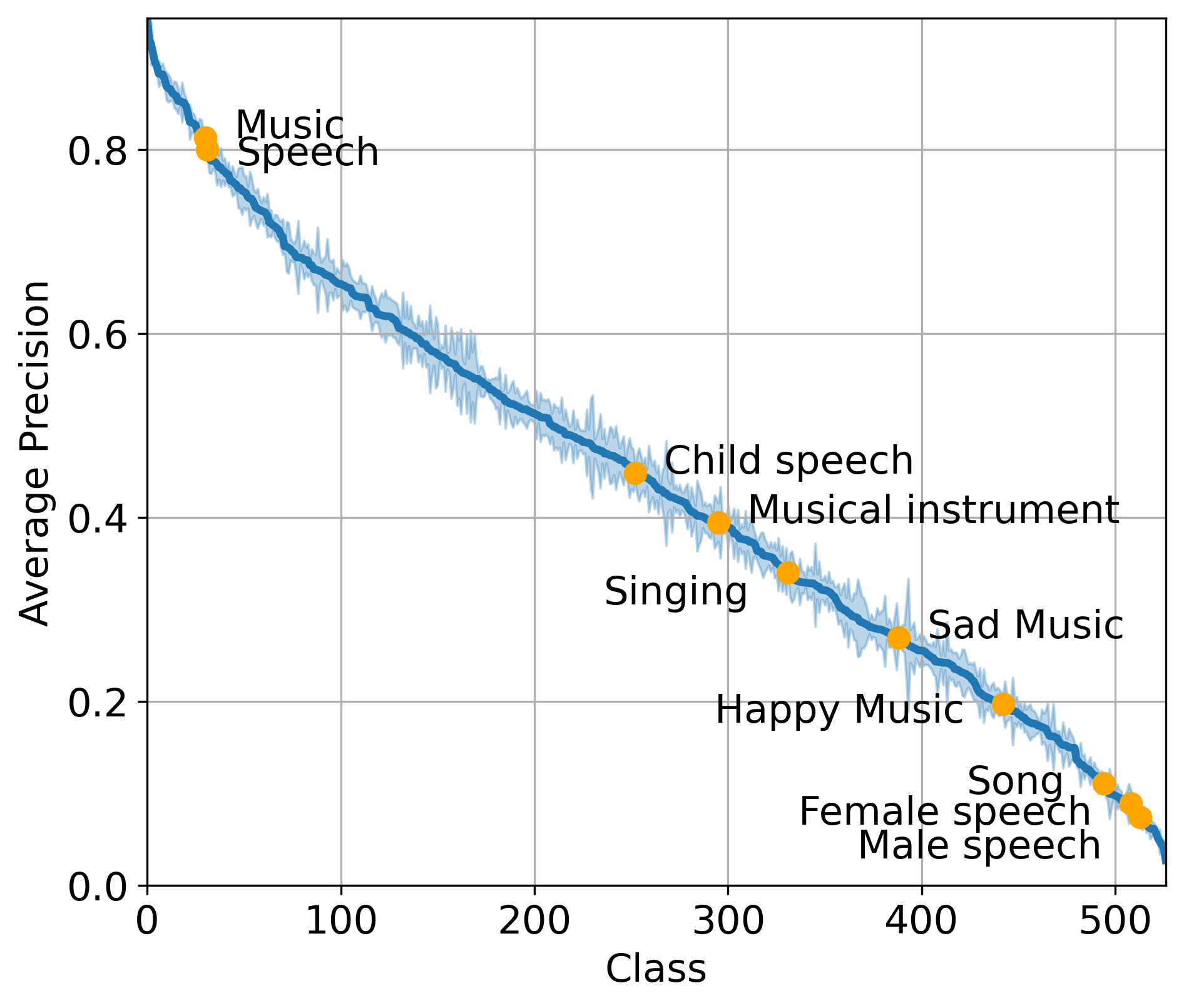}
  \caption{Sorted class-wise average precision (AP) and its standard deviation of the model trained on full set. Note that the ``Speech'' class has a much higher AP than the ``Male Speech'', ``Female Speech'', and ``Child Speech'' class. Similarly, the ``Music'' class has a much higher AP than the ``Happy Music'' and ``Sad Music'' class. ``Singing'' and ``Song'' have similar definition but very different AP. Classes with low AP also have a larger AP variance.}
  \label{fig:classAP}
\end{figure}

\section{Label Enhancement}
\label{sec:label}

In this section, we explore the noisy label aspect of AudioSet: how it impacts audio tagging performance, and how to alleviate it. This line of research is motivated by observing the model's class-wise performance. In Figure~\ref{fig:classAP}, we show the class-wise average precision (AP) of the model trained with the full set. From the figure it is immediately apparent that the AP of each class differs greatly, indicating that the model has a range of ability to recognize various sounds. This is not an issue specific to our model or training pipeline, but has been widely reported in prior work~\cite{kong2020panns,kong2019weakly,fonseca2020addressing,shah2018closer,ford2019large}. The order of class-specific performance reported by independent research also appears to be similar. For example, the ``Male speech'', ``Bicycle'', ``Harmonic'', ``Rattle'', and ``Scrape'' classes are among the 10 worst performing classes in~\cite{shah2018closer}, and they are also are among the 10 worst performing classes for our model when trained with the balanced set. We further confirm that models with different architectures have similar class-specific performance order with experiments in Section~\ref{sec:various_model}. This consistency suggests that the issue might be due to an intrinsic problem with the data or the task.  Since the class-wise AP is not strongly correlated with either class sample count in the training set or the class annotation quality estimate released by the AudioSet authors (as shown in Table~\ref{tab:corr}), it has been hypothesized that the class-wise performance variation is due in part to the difficulty in reliably tagging the different sound classes themselves~\cite{kong2020panns,ford2019large}.

\begin{table}
\centering
\caption{Correlation Coefficients Between Class-wise AP and Class Sample Count/Annotation Quality Estimate Released by AudioSet Authors.}
\label{tab:corr}
\begin{tabular}{@{}lcc@{}}
\toprule
                                                       & Balanced Set & Full Set \\ \midrule
AP and Sample Count                & 0.1692       & 0.0946   \\
AP and Annotation Quality Estimate & 0.2464       & 0.2629   \\ \bottomrule
\end{tabular}
\end{table}

While we agree that the poor performance of some classes could be due to particular audio events being difficult to identify, it is not true for all poor-performing classes. For example, the ``Male Speech'', ``Female Speech'', and ``Child Speech'' classes have APs of 0.07, 0.09, 0.45, respectively while the AP of the ``Speech'' class is 0.80. This discrepancy cannot be explained by the class difficulty hypothesis because recognizing speaker gender from speech is a relatively easy task~\cite{wu1991gender,childers1991gender,wang2017learning}, and the performances of the speech classes should not be so disparate. By examining the class sample counts, we find another issue that the sample count of the ``Speech'' class is substantially larger than the sum of sample counts of the ``Male Speech'', ``Female Speech'', and ``Child Speech'' classes. Specifically, in the balanced set, there are 5,309 audio clips with the label ``Speech'' but only 55, 55, 128 audio clips are with label ``Male Speech'', ``Female Speech'', and ``Child Speech'', respectively. The same thing happens in the full set (shown in Figure~\ref{fig:samplecnt}): the ``Speech Class'' has 947,009 samples while the sum of the other three classes is 34,878. In other words, only 4.5\% and 3.7\% of speech samples are labeled as either male, female, or child speech in the balanced and full AudioSet, respectively. This indicates that a large portion of samples are not correctly labeled. Based on these two observations, we hypothesize that the low performance of the male, female, and child speech classes is not due a small number of samples, or inherent classification difficulty, but that they have only a small fraction of correctly labeled data, which ultimately confuses the model. We refer to this phenomenon as a Type I error.

We also find that there are substantial samples labeled with sub-classes, but not with the corresponding parent class defined by the AudioSet ontology. For example, there are 40 and 3,201 audio clips labeled as either ``Male Speech'', ``Female Speech'', or ``Child Speech'', but not labeled as ``Speech'' in the balanced and full AudioSet, respectively. We refer to this phenomenon as Type II error.

We formalize the two types of error as follows: 
\begin{enumerate}
    \item Type I error: an audio clip is labeled with a parent class, but not also labeled as a child class when it does in fact contain the audio event of the child class.
    \item Type II error: an audio clip is labeled with a child class, but not labeled with corresponding parent classes.
\end{enumerate}

It is worth mentioning that neither type of error are  included in the quality estimate released by the AudioSet authors because the quality estimate checked 10 random audio clips of each class and verified that they actually contained the corresponding sound event. In other words, the quality estimate counts the false positive annotation errors, but not false negatives. As a consequence, the quality estimate of the ``Male Speech'', ``Female Speech'', and ``Child Speech'' is 90\%, 100\%, and 100\%, respectively, while they have obvious false negative annotation errors. 

Unfortunately, false negatives are prevalent in AudioSet. Another example are the music classes (see Figure~\ref{fig:samplecnt} and~\ref{fig:classAP} for sample counts and class-wise AP of music classes). The reason for these types of errors is due to the AudioSet annotation pipeline. In the pipeline, the human annotator verifies the candidate labels nominated by a series of automatic methods (e.g., by using metadata). Also, the list of candidate labels is limited to ten labels per clip. Since the automatic methods for nomination are not perfect, some existing sound events fail to be nominated, or are nominated but ranked below the top ten, thus leading to missing labels~\cite{fonseca2020addressing,gemmeke2017audio}. 

As seen in the speech class example, annotation error can impact performance, but has not received much attention. To the best of our knowledge, only a few efforts have covered the missing label issue. In~\cite{shah2018closer,meire2019impact}, a synthetic error is studied, however, the real-world noisy labels are believed to be much harder to deal with than the synthetic labels. In~\cite{fonseca2020addressing}, the authors propose a loss masking based teacher-student model. In this section, we propose an ontology-based label enhancement method to alleviate the noisy label problem. Our approach differs from previous work in three aspects: First, we work on real-world noisy labels rather than synthetic corrupted labels; Second, we explicitly modify the labels of the training data rather than using loss masking during training. Thus the enhanced label set can be used in the exact same way as the original set (no need to modify the model and training pipeline). We plan to release the enhanced label set to facilitate future research.  Third, we leverage the AudioSet ontology to constrain label modification, which reduces the chance of incorrect modifications. For example, for an audio clip labeled as ``Speech'', we only consider adding child or parent labels in the specific "Speech" branch of the ontology.

\begin{algorithm}[!t]
\caption{Label Enhancement}\label{alg:2}
\begin{algorithmic}[1]
\Require{
\quad \newline
Teacher Model $\mathcal{M}$ \newline
Dataset $\mathcal{D}=\{\boldsymbol{x}^{(i)}, \boldsymbol{y}^{(i)}\}$, $i \in \{1,...,N\}$ \newline
Label Ontology $\mathcal{O}$}
\Algphase{Procedure 1: Generate Label Modification Threshold}
\textbf{Input:} $\mathcal{M}, \mathcal{D}$ \newline
\textbf{Output:} Threshold Set $\mathcal{T}=\{\boldsymbol{t}_{k}\}, k \in \{1,...,527\}$
\For{$k\in \{1,...,527\}$}
\State {$t_k=\sum_{i=1}^{N}\mathbbm{1}_{\{k\in\boldsymbol{y}^{(i)}\}}\mathcal{M}(\boldsymbol{x}^{(i)})(k)/\sum_{i=1}^N\mathbbm{1}_{\{k\in\boldsymbol{y}^{(i)}\}}$ }
\EndFor
\Return $\mathcal{T}=\{\boldsymbol{t}_{k}\}$

\Algphase{Procedure 2: Enhance the Label Set}
\textbf{Input:} $\mathcal{M}, \mathcal{D}, \mathcal{O}, \mathcal{T}$ \newline
\textbf{Output:} Enhanced Label Set $\{\boldsymbol{y}^{\prime(i)}\}$, $i \in \{1,...,N\}$
\State Initialize $\{\boldsymbol{y}^{\prime(i)}\}=\{\boldsymbol{y}^{(i)}\}$
\For{$i \in \{1,...,N\}$ }
\For{$k \in \boldsymbol{y}^{(i)}$ }
\For{$k_n \in \mathcal{O}(k)$} \Comment{parent or child class of $k$}
\If{$\mathcal{M}(\boldsymbol{x}^{(i)})(k_n)>t_{k_n}$ and $k_n\not\in\boldsymbol{y}^{(i)}$}
\State $\boldsymbol{y}^{\prime(i)}=\boldsymbol{y}^{\prime(i)}\cup\{k_n\}$
\EndIf
\EndFor
\EndFor
\EndFor
\Return $\{\boldsymbol{y}^{\prime(i)}\}$
\end{algorithmic}
\end{algorithm}

As shown in Algorithm~\ref{alg:2}, the proposed approach consists of the following steps. First, we train a teacher model using the full AudioSet with the original label set. Second, we set a label modification threshold for each audio tagging, specifically, we set the threshold of a class as the teacher model's mean prediction score of all audio clips originally labeled as that class (lines 1-2). The threshold can also be set as other values such as the 5th, 10th, or 25th percentile of the teacher model's prediction score. The lower the threshold, the more labels are added. We then identify all samples that need to be relabeled. For each sample, we compile all child (Type I) and/or parent (Type II) labels of all original labels as the candidate set according to the AudioSet ontology (line 6). For each label in the candidate set, if the teacher model's prediction score of the class is greater than the corresponding label modification threshold, we add it to the labels of the sample (line 7-8). Finally, we retrain the model from scratch with the enhanced label set. 

\begin{table}[t]
\centering
\caption{Result of Label Enhancement on the Balanced Set (Note the mAP without Label Enhancement is 0.3108$\pm$0.0013).}
\label{tab:labelres}
\begin{tabular}{@{}lccc@{}}
\toprule
                          & Type I & Type II & Type I and II \\ \midrule
\# Impact Classes           & 212     & 93     & 274           \\
Label Added (\%)          & 3.7\%  & 3.9\%   & 7.2\%         \\ \midrule
Impacted Class Improvement      & 4.5\%  & 3.8\%   & 4.5\%         \\
Non-impacted Class Improvement  & 1.9\%  & 2.1\%   & 1.3\%         \\
Mean Class-wise Relative Improv. & 3.0\%  & 2.4\%   & 2.9\%         \\ \midrule
mAP Improvement           & 1.9\%  & 1.5\%   & 1.7\%         \\ 
mAP                       & \begin{tabular}[c]{@{}c@{}}0.3166\\ $\pm$0.0016\end{tabular} & \begin{tabular}[c]{@{}c@{}}0.3156\\ $\pm$0.0007\end{tabular}  & \begin{tabular}[c]{@{}c@{}}0.3162\\ $\pm$0.0005\end{tabular}        \\
\bottomrule
\end{tabular}
\end{table}

\begin{table*}[h]
\centering
\caption{AudioSet label enhancement (LE) experiment results. We use the mean, 25th percentile (25P), 10th percentile (10P), and 5th percentile (5P) of the prediction score as the label modification thresholds and generate 4 enhanced AudioSet training label sets and evaluation label sets. We then train the model with the enhanced training sets and evaluate it on various evaluation sets. The results show that the model trained with enhanced label sets consistently outperforms the model trained with original label sets on all evaluation sets except the original AudioSet evaluation set.}
\label{tab:labelres2}
\begin{tabular}{@{}lcccccccc@{}}
\toprule
         & \begin{tabular}[c]{@{}c@{}}Label \\ Added (\%)\end{tabular} & \begin{tabular}[c]{@{}c@{}}ESC-50\\ 40 Classes\end{tabular} & \begin{tabular}[c]{@{}c@{}}FSD50k\\ Eval\end{tabular} & \begin{tabular}[c]{@{}c@{}}AudioSet\\ Eval Ori\end{tabular} & \begin{tabular}[c]{@{}c@{}}AudioSet \\ Eval Mean\end{tabular} & \begin{tabular}[c]{@{}c@{}}AudioSet \\ Eval 25P\end{tabular} & \begin{tabular}[c]{@{}c@{}}AudioSet \\ Eval 10P\end{tabular} & \begin{tabular}[c]{@{}c@{}}AudioSet \\ Eval 5P\end{tabular} \\ \midrule
\multicolumn{9}{l}{AudioSet Balanced Training Set}                                                              
\\ \midrule
No LE    & 0.0                                                         & 0.7320                                                      & 0.3443                                                & 0.3123                                                      & 0.3485                                                        & 0.3632                                                       & 0.3540                                                       & 0.3417                                                      \\
LE, Mean & 7.2                                                         & 0.7573                                                      & 0.3549                                                & 0.3162                                                      & 0.3591                                                        & 0.3739                                                       & 0.3630                                                       & 0.3500                                                      \\
LE, 25P  & 22.8                                                        & \textbf{0.7639}                                             & 0.3680                                                & \textbf{0.3165}                                             & \textbf{0.3632}                                               & 0.3855                                                       & 0.3760                                                       & 0.3628                                                      \\
LE, 10P  & 44.5                                                        & 0.7551                                                      & 0.3639                                                & 0.3078                                                      & 0.3527                                                        & 0.3840                                                       & 0.3811                                                       & 0.3699                                                      \\
LE, 5P   & 60.2                                                        & 0.7548                                                      & \textbf{0.3766}                                       & 0.3078                                                      & 0.3518                                                        & \textbf{0.3862}                                              & \textbf{0.3880}                                              & \textbf{0.3790}                                             \\ \midrule
\multicolumn{9}{l}{AudioSet Full Training Set}  
\\ \midrule
No LE    & 0.0                                                         & 0.8587                                                      & 0.4977                                                & \textbf{0.4397}                                             & 0.5053                                                        & 0.5143                                                       & 0.4930                                                       & 0.4723                                                      \\
LE, Mean & 11.1                                                        & \textbf{0.8772}                                             & 0.5079                                                & 0.4386                                                      & \textbf{0.5075}                                               & 0.5190                                                       & 0.4977                                                       & 0.4769                                                      \\
LE, 25P  & 37.3                                                        & 0.8736                                                      & \textbf{0.5097}                                       & 0.4296                                                      & 0.4999                                                        & \textbf{0.5267}                                              & 0.5093                                                       & 0.4891                                                      \\
LE, 10P  & 77.7                                                        & 0.8608                                                      & 0.5078                                                & 0.4094                                                      & 0.4752                                                        & 0.5178                                                       & \textbf{0.5121}                                              & 0.4969                                                      \\
LE, 5P   & 111.9                                                       & 0.8534                                                      & 0.4988                                                & 0.3936                                                      & 0.4560                                                        & 0.5047                                                       & 0.5088                                                       & \textbf{0.4987}                                             \\ \bottomrule
\end{tabular}
\end{table*}

We apply the proposed label enhancement method (with the teacher model's mean prediction score as the label modification threshold) on the balanced training set and show the results in Table~\ref{tab:labelres}. Note the model without label enhancement has an mAP of 0.3108$\pm$0.0013 (the model from the previous section). The key findings are as follows: First, a noticeable number of labels are added, and over half of the classes are impacted, which further indicates that the missing label issue is prevalent in AudioSet. Second, enhancing the label improves the performance of both impacted and non-impacted classes, but the impacted classes have a larger relative improvement. Third, the mean class-wise relative AP improvement is larger than the relative mean AP (mAP) improvement, indicating that more of the classes that improved originally had below-average performance. This supports our hypothesis that the missing label problem lowers the performance of a sound class. Fourth, we evaluate the performance of fixing Type I errors, Type II errors, and fixing both. The improvement achieved by fixing Type I errors is larger than fixing Type II errors. Fixing both cannot further improve the performance. Fifth, since the performance improvement is relatively minor, we run all experiments three times with different random seeds and report both the mean and standard deviation. As shown in the table, the results verify the statistical significance of the improvement. Finally, we also applied the label enhancement method on the full AudioSet, however, we did not observe a performance improvement. Fixing Type I, Type II, and both errors leads to mAPs of 0.4400, 0.4387, and 0.4386, respectively, while the model without label enhancement achieves an mAP of 0.4397$\pm$0.0007. We believe the main reason for the relatively small improvement achieved by label enhancement is that the same label noise exists consistently in both the training set and evaluation set. Therefore, merely applying label enhancement on the training set leads to a mismatch between the training and evaluation sets.  The performance results do not therefore fully reflect the actual improvement. In addition, it is possible that the label modification threshold is not appropriate for the full AudioSet.

In order to verify these hypotheses, we evaluate our model on existing datasets with more accurate annotation including ESC-50~\cite{piczak2015esc} and FSD50K~\cite{fonseca2020fsd50k}, and also test various label modification thresholds. ESC-50 contains 2,000 audio samples of 50 sound classes, among which 40 classes are overlapped with the AudioSet. Therefore, we evaluate our model trained with AudioSet on the 1,600 samples that are labeled as these 40 overlapped classes. FSD50K is a recently collected data set of sound event audio clips with 200 classes drawn from the AudioSet ontology. The FSD50K evaluation set is more carefully annotated compared with the training and validation set and can be used as fair references. Since the length of AudioSet model input is 10s while a small portion of FSD50K audio clips are longer than 10s, we cut all FSD50K audio clips to 10s for testing. In addition, we also apply the proposed label enhancement algorithm on the AudioSet evaluation set and generate enhanced evaluation sets. We include the enhanced AudioSet evaluation sets as additional evaluation sets.

We evaluate various label modification thresholds including the mean, 25th percentile (25P), 10th percentile (10P), and 5th percentile (5P) of the teacher model's prediction score of all audio clips originally labeled as that class. The lower the threshold, the more labels are modified, e.g., using the 5th percentile of the prediction score as the threshold changes the largest number of labels. We then train models with the four enhanced label sets and compare their results on seven evaluation sets (ESC-50, FSD50K, original AudioSet evaluation set, and four enhanced AudioSet evaluation set with different label modification thresholds).

As shown in Table~\ref{tab:labelres2}, we find that models trained with enhanced AudioSet label sets consistently outperforms the model trained with the original AudioSet label set on all evaluation sets except the original AudioSet evaluation set, demonstrating that the proposed label enhancement algorithm is able to improves the model performance, the reason why we cannot observe the improvement on the AudioSet evaluation set is that the evaluation set itself contains annotation errors. While there is no threshold that is optimal for all evaluation sets, for both balanced and full AudioSet experiments, we find the mean and 25th percentile of the teacher model's prediction score are the most appropriate label modification thresholds.

We believe it is an important and non-negligible topic for future AudioSet and general audio tagging research because noisy labels are inevitable for a large-scale dataset and errors will impact model performance. In the following section, we use models trained with the enhanced label set as default for all balanced set experiments.

\section{Weight Averaging and Ensemble}
\label{sec:ensemble}

\subsection{Model Weight Averaging}

In this section, we explore improving model performance by aggregating multiple models. The first strategy we explore is \emph{weight averaging}~\cite{izmailov2018averaging}. Weight averaging performs an equal average of the weights traversed by the optimizer, which makes the solution fall in the center, rather than the boundary, of a wide flat low-loss region and thus lead to better generalization than conventional training. Empirically, weight averaging has been shown to improve the performance of various models such as VGG~\cite{simonyan2014very}, ResNets~\cite{he2016deep}, and DenseNets~\cite{huang2017densely} on a variety of tasks~\cite{izmailov2018averaging,athiwaratkun2019there}. While weight averaging is usually applied with a high constant or cyclical learning rate, we find it is helpful even when used together with a weight decay strategy. 

\begin{table}[t]
\centering
\caption{Performance Impact on mAP Due to Weight Averaging}
\label{tab:wa}
\begin{tabular}{@{}lcc@{}}
\toprule
\multicolumn{1}{c}{}   & Balanced Set   & Full Set        \\ \midrule
Without Weight Averaging & 0.3162 $\pm$ 0.0005 & 0.4397 $\pm$ 0.0007 \\
With Weight Averaging    & 0.3192 $\pm$ 0.0015 & 0.4435 $\pm$ 0.0008 \\ \bottomrule
\end{tabular}
\end{table}

\begin{figure}[t]
  \centering
  \includegraphics[width=8.5cm]{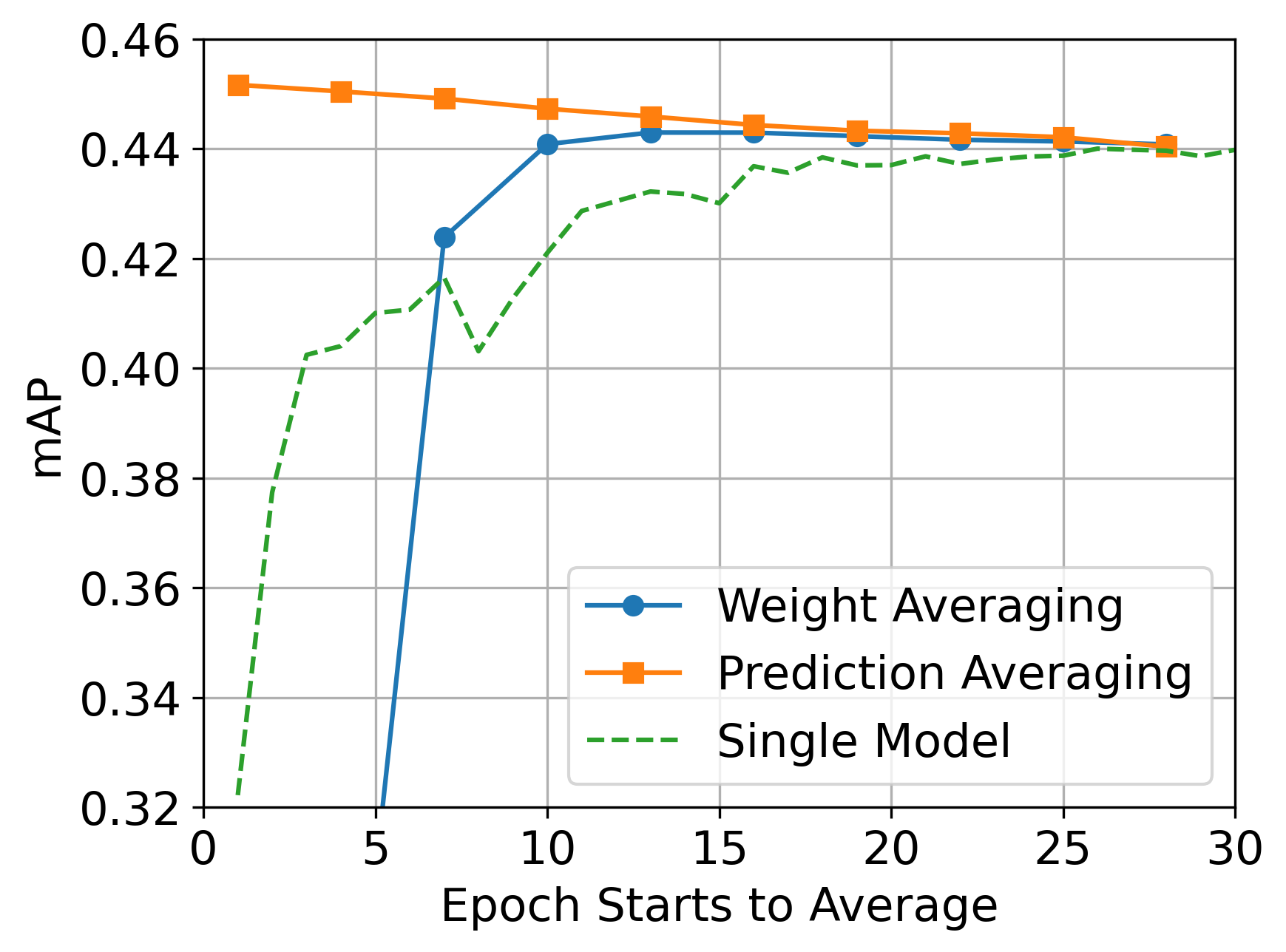}
  \caption{Relationship of the performance of averaging models with the epoch starts to average. For both weight and prediction averaging, we average all checkpoints from the starting epoch to the last epoch, i.e., the earlier to start averaging, the more checkpoints are averaged. Note that the improvement of model averaging is not sensitive to exactly when weight averaging begins. For weight averaging, the optimal starting epoch is around the $15^{th}$ epoch while starting averaging at any epoch after the $10^{th}$ epochs can outperform any single checkpoint. For prediction averaging, starting averaging from the first epoch leads to the highest mAP, indicating averaging all checkpoints is optimal, while starting averaging at any epoch can outperform any single checkpoint. However, averaging the predictions of the last few checkpoints barely outperforms single checkpoints, indicating the importance of diversity.}
  \label{fig:ensemble}
\end{figure}

In this work, we simply average all weights of the model checkpoints at multiple epochs. For both balanced set and full set experiments, we start averaging model checkpoints of every epoch after the learning rate is decreased to 1/4 of the initial learning rate (i.e., the $41^{st}$ and the $16^{th}$ epochs, respectively) until the end of the training. As shown in Table~\ref{tab:wa}, weight averaging leads to a 0.9\% improvement for both balanced set and full set experiment. We further find the improvement is not sensitive to exactly when weight averaging begins.  As shown in Figure~\ref{fig:ensemble}, starting averaging at any epoch after the $10^{th}$ epochs (until the last epoch) can outperform any single checkpoint model for the full set experiment. 

In summary, weight averaging is easy to implement, adds no additional cost to training and inference, but can consistently improve model performance. By applying weight averaging to our models, we get our best single model with an mAP of 0.3192 and 0.4435 for balanced and full AudioSet experiment, respectively.

\subsection{Ensemble}

Finally, we explore a series of ensemble strategies. The goal of ensemble methods is to combine the predictions of several models to improve generalizability and robustness over any single model. Previously, ensemble of audio tagging models has been studied in in~\cite{lee2017ensemble,shi2019hodgepodge,guo2019multi,lim2019sound,lopez2020ensemble,palanisamy2020rethinking,ford2019deep}, but typically only one strategy is covered in each of these previous efforts. In this work, we use the simple voting algorithm, but compare multiple ways of building the model committee. The reason why we do not use iterative ensemble methods (e.g., Boosting) is because AudioSet training is  expensive making iterative training computationally unreasonable for this work.

\begin{table}[]
\centering
\caption{Results of Model Ensemble. For each experiment, we show the number of the models in the committee (\# Models), the average mAP of models in the committee (Avg mAP), the mAP of the best model in the committee (Best mAP), and the mAP of the ensemble model (Ensemble mAp). Note that for all experiments, the ensemble mAP is higher than the best mAP.}
\label{tab:ensemble}
\begin{tabular}{@{}lcccc@{}}
\toprule
\multicolumn{1}{c}{}  & \# Models  & Avg mAP  & Best mAP & Ensemble mAP \\ \midrule
\multicolumn{5}{l}{Checkpoints of a Single Run}            \\ \midrule
Balanced              & 60        & 0.2369    & 0.3169   & 0.3280    \\
Full                  & 30        & 0.4236    & 0.4406   & 0.4518    \\ \midrule
\multicolumn{5}{l}{Multiple Runs with Same Setting}        \\ \midrule
Balanced              & 3         & 0.3162    & 0.3167   & 0.3446    \\
Full                  & 3         & 0.4397    & 0.4405   & 0.4641    \\ \midrule
\multicolumn{5}{l}{Models Trained with Different Settings} \\ \midrule
Bal-pretrain          & 2         & 0.1978    & 0.2385   & 0.2410    \\
Bal-mixup rate        & 5         & 0.3009    & 0.3123   & 0.3476    \\
Bal-mixup-$\alpha$    & 3         & 0.3071    & 0.3123   & 0.3418    \\
Bal-augment           & 3         & 0.2775    & 0.3123   & 0.3281    \\
Bal-label           & 4         & 0.3146    & 0.3169   & 0.3503    \\
Bal-top5              & 5         & 0.3168    & 0.3180   & 0.3527    \\
Bal-all               & 20        & 0.2987    & 0.3180   & {\bf 0.3620}    \\ \cmidrule(r){1-1}
Full-pretrain         & 2         & 0.3831    & 0.3939   & 0.4006    \\
Full-augment          & 4         & 0.4080    & 0.4396   & 0.4578    \\
Full-label    & 4         & 0.4397    & 0.4400   & 0.4653    \\
Full-top5             & 5         & 0.4396    & 0.4405   & 0.4690    \\
Full-all              & 10        & 0.4201    & 0.4405   & {\bf 0.4744}    \\ \bottomrule
\end{tabular}
\end{table}

\subsubsection{Checkpoint Averaging}

The first strategy investigated is checkpoint averaging, whereby the output of checkpoint models at multiple epochs are averaged together. The implementation is similar to weight averaging, but is conducted in the model space rather than the weight space. Since we conduct random sampling with replacement during full set training, the combination with checkpoint averaging is the same as bootstrap aggregating (i.e., Bagging)~\cite{breiman1996bagging}. In our experiment, we average the output of all checkpoint models (i.e., 60 and 30 checkpoint models for the balanced set and full set, respectively). As shown in the upper part of Table~\ref{tab:ensemble}, this approach works well. Specifically, the ensembled model noticeably outperforms the best checkpoint model in the committee. In addition, as shown in Figure~\ref{fig:ensemble}, starting averaging from the first epoch leads to the highest mAP, indicating averaging all checkpoints is optimal. Averaging from any epoch can outperform the best single checkpoint model, which can be a simple alternative. However, this approach greatly increases the computational overhead of inference, which makes it less practical in deployment.

\subsubsection{Averaging Models Trained with Different Random Seeds}

Previous work suggests that ensembles generalize better when they constitute members that form a \emph{diverse} and \emph{accurate} set~\cite{chandra2006trade}. As shown in Figure~\ref{fig:ensemble}, starting averaging the checkpoint predictions from the last few epochs can only slightly outperform the best single checkpoint model, even though these checkpoint models are quite accurate, indicating the importance of diversity. Therefore, we run the experiment three times with the exact same setting, but with a different random seed. We then average the output of the last checkpoint model of each run. As shown in the middle part of Table~\ref{tab:ensemble}, this approach leads to an even larger improvement than checkpoint averaging with only three models in the committee. Therefore, averaging models trained with different random seeds, while increasing the training cost (due to the repeat runs), is more practical for deployment and offers better performance.

\subsubsection{Averaging Models Trained with Different Settings}
Finally, we explore averaging more models with greater diversity. Specifically, we ensemble models trained with all different settings tested in this paper, including whether pretraining is used (pretrain), different mix-up rates (mixup rate), different mix-up $\alpha$ (mix-up-$\alpha$), different augmentation settings (augment), and different label enhancement strategies (label). As shown in the lower part of Table~\ref{tab:ensemble}, no matter how the model committee is built, ensemble always improves the performance and outperforms the best model in the committee. In the literature, diversity is usually introduced with an intuitive motivation.  For example, in~\cite{lee2017ensemble}, the authors ensemble models use different scale inputs because they believe the optimal input scale varies with the target audio events, and ensembles allows the model to extract relevant information from inputs with various scales. But according to our experimental results, the source of the diversity seems to be less important, i.e., the diversity caused by any factor is helpful for an ensemble.

In addition, we find the performance of the ensemble model is positively correlated with the accuracy of the models in the committee as well as the number of the models. For both the balanced set and full set experiments, our best model is achieved when all available models form an ensemble. 

\section{Supplementary Experiments}
\label{sec:ablation}
\begin{table}[h]
\centering
\caption{Ablation study results on AudioSet.}
\label{tab:ablas}
\begin{tabular}{@{}lcc@{}}
\toprule
                               & Balanced AudioSet & Full AudioSet \\ \midrule
PSLA Model                     & \bf{0.3280}            & \bf{0.4518}        \\
PSLA Model - Pretrain          & 0.2379            & 0.4302        \\
PSLA Model - Balanced Sampling & -                 & 0.3688        \\
PSLA Model - Masking           & 0.3154            & 0.4430        \\
PSLA Model - Mixup             & 0.3181            & 0.4493        \\
PSLA Model - Label Enhancement & 0.3229            & -             \\
PSLA Model - Ensemble          & 0.3162            & 0.4397        \\
PSLA Model - Ensemble + WA     & 0.3192            & 0.4435        \\ \bottomrule
\end{tabular}
\end{table}

\subsection{Ablation Study}

From Section~\ref{sec:pretrain} to Section~\ref{sec:ensemble}, we incrementally improve model performance from the baseline by incorporating a new technique with other techniques that have been found to be effective. In order to clearly identify the contribution of each technique and verify that all are necessary for the best model, we conduct an ablation study on balanced and full AudioSet. Specifically, we set the PSLA model with checkpoints ensemble as the baseline (the best model for a single training run), and then remove techniques from PSLA one by one, and check the performance. As shown in Table~\ref{tab:ablas}, removing any technique from PSLA leads to a performance drop, demonstrating that all proposed techniques are useful. It is worth mentioning that removing balanced sampling leads to a significant performance drop for AudioSet, the performance of the model is worse than the model only with pretraining (0.3939 mAP, in Table~\ref{tab:aug}), indicating that other techniques (e.g., masking, mixup, and ensemble) should be used together with balanced sampling for AudioSet. Besides balanced sampling, removing pretraining leads to the largest performance drop, followed by ensemble, time and frequency masking, and mixup training for the full AudioSet. 

\begin{table*}[h]
\centering
\caption{Comparison of the performance on mAP of various models trained with PSLA and without PSLA on the balanced and full AudioSet.}
\label{tab:moremodel}
\begin{tabular}{@{}cccccccc@{}}
\toprule
\multicolumn{1}{l}{}                                                                & \multicolumn{1}{l}{\multirow{2}{*}{\# Params}} & \multicolumn{3}{c}{Balanced AudioSet} & \multicolumn{3}{c}{Full AudioSet}                                                      \\ \cmidrule(l){3-8} 
                                                                                    & \multicolumn{1}{l}{}                           & No PSLA     & PSLA      & Imp.(\%)    & No PSLA & PSLA                                                              & Imp.(\%) \\ \midrule
MobileNet V2                                                                            & 2.90M                                          & 0.1612      & 0.2650    & 64.4        & 0.3032  &      \begin{tabular}[c]{@{}c@{}}0.4058\\ (Single: 0.3940)\end{tabular}                                                        & 33.8     \\
EfficientNet-B0, Single-headed Attention  & 5.36M                                          & 0.1529      & 0.3350    & 119.1       & 0.3789  &       \begin{tabular}[c]{@{}c@{}}0.4493\\ (Single: 0.4391)\end{tabular}                                                      & 18.6     \\
EfficientNet-B2, Mean Pooling            & 8.44M                                          & 0.1903      & 0.3317    & 74.3        & 0.3325  &  \begin{tabular}[c]{@{}c@{}}0.4455\\ (Single: 0.4382)\end{tabular}                                                           & 34.0     \\
EfficientNet-B2, Single-headed Attention & 9.19M                                          & 0.1478      & 0.3406    & 130.4       & 0.3818  & \begin{tabular}[c]{@{}c@{}}0.4556\\ (Single: 0.4414)\end{tabular} & 19.3     \\
EfficientNet-B2, 4-headed Attention     & 13.64M                                         & 0.1570      & 0.3280    & 108.9       & 0.3723  & \begin{tabular}[c]{@{}c@{}}0.4518\\ (Single: 0.4435)\end{tabular} & 21.4     \\
ResNet-50, Single-headed Attention                                                                              & 25.66M                                         & 0.1635      & 0.3180    & 94.5        & 0.3790  &   \begin{tabular}[c]{@{}c@{}}0.4477\\ (Single: 0.4042)\end{tabular}                                                          & 18.1     \\ \bottomrule
\end{tabular}
\end{table*}

\subsection{Experiment with Various Audio Tagging Models}
\label{sec:various_model}
In the previous sections, we focus on the EfficientNet-B2 with a 4-headed attention model described in Section~\ref{sec:base_mdl}. In order to identify if the proposed PSLA framework is model-agnostic and explore the model size-performance trade-offs, in this section, we evaluate the PSLA framework using 6 different models. All models take the same input and are trained with the same setting as mentioned in Section~\ref{sec:train}.
\begin{enumerate}
    \item MobileNet V2~\cite{sandler2018mobilenetv2}. The MobileNet model does not have an attention module. We use a fully connected layer as the classification layer.
    \item EfficientNet-B0 with single-headed attention model. The model architecture is the same as the model described in Section~\ref{sec:base_mdl} except that it is based on a smaller EfficientNet-B0 and only has one attention module.
    \item EfficientNet-B2 with mean pooling model. The model architecture is the same as the model described in Section~\ref{sec:base_mdl} except that it uses mean pooling rather than attention pooling.
    \item EfficientNet-B2 with single-headed attention model. The model architecture is the same as the model described in Section~\ref{sec:base_mdl} except that it only has one attention module.
    \item EfficientNet-B2 with 4-headed attention model. This is the model we use in from Section~\ref{sec:pretrain} to Section~\ref{sec:ensemble} and is described in Section~\ref{sec:base_mdl}.
    \item ResNet50 with single-headed attention module. This is the model proposed in~\cite{ford2019deep}.
\end{enumerate}

To save compute, for all PSLA models, we use the checkpoint averaging ensemble that only requires a single training process, we also report the single model with weight averaging for all full AudioSet experiments. As shown in Table~\ref{tab:moremodel}, when trained with PSLA techniques, all models can achieve a noticeable performance improvement. This justifies that the proposed PSLA framework is model-agnostic. 

Comparing the EfficientNet-B2 models with 4-headed attention, single-headed attention, and mean pooling, we find while the single 4-headed attention model performs best (0.4435 mAP), the single-headed attention model and the mean pooling model only perform slightly worse.
The EfficientNet-B0 model with single-headed attention that has 5.36M parameters also achieves a comparable performance with the best existing model that has 81M parameters~\cite{kong2020panns}. The choice of the model depends on the application, e.g., attention-based models can be used for frame-level tagging; models with mean pooling can be used for streaming applications; smaller models are preferable for resource-constrained devices. 

We also compute the Pearson correlation of class-wise APs between these models and find that the correlation of class-wise APs are high (over 0.95), this confirms that the poor performance of some class is not due to model architecture, but due to the data.

\begin{table}[h]
\centering
\caption{Experiment result on FSD50K dataset.}
\label{tab:ablfsd}
\begin{tabular}{lc}
\toprule
                               & FSD50K Eval \\ \midrule
FSD50K Baseline~\cite{fonseca2020fsd50k}    & 0.434      \\
Audio Transformers~\cite{verma2021audio}    & 0.537      \\
PSLA Model                     & \bf{0.5671}      \\  \midrule
PSLA Model - Pretrain          & 0.4524      \\
PSLA Model - Balanced Sampling & 0.5626      \\
PSLA Model - Masking           & 0.5617      \\
PSLA Model - Mixup             & 0.5164      \\
PSLA Model - Label Enhancement & 0.5583      \\
PSLA Model - Ensemble          & 0.5535      \\
PSLA Model - Ensemble + WA     & 0.5571      \\ \bottomrule
\end{tabular}
\end{table}

\subsection{Experiment on FSD50K}

In the previous sections, we focus on AudioSet. To check the generalizability of the proposed PSLA techniques, we also conduct a set of experiments on FSD50K~\cite{fonseca2020fsd50k}. Specifically, we train the EfficientNet-B2 model with a 4-headed attention module with an initial learning rate of 5e-4 and a batch size of 24 for 40 epochs. The learning rate is cut in half every 5 epochs after the $10^{th}$ epoch. Since the maximum input audio length of FSD50K is 30s, we pad all input audio clips to 30s. For the single model, we train it with the FSD50K training set, validate it on the FSD50K validation set, and evaluate it on the FSD50K evaluation set. We use the same weight averaging and checkpoint averaging ensemble setting as the AudioSet experiments. We also conduct an ablation study on FSD50K. 

As shown in Table~\ref{tab:ablfsd}, our single model, weight averaging model, and ensemble model achieve an mAP of 0.5535, 0.5571, and 0.5671 on the FSD50K evaluation set, respectively, all outperform the best existing model~\cite{verma2021audio}. Removing any technique from PSLA leads to a performance drop, demonstrating that all proposed techniques can be generalized to the FSD50K dataset.

\subsection{Learning Curve of PSLA models}

\begin{figure}[h]
  \centering
  \includegraphics[width=8.5cm]{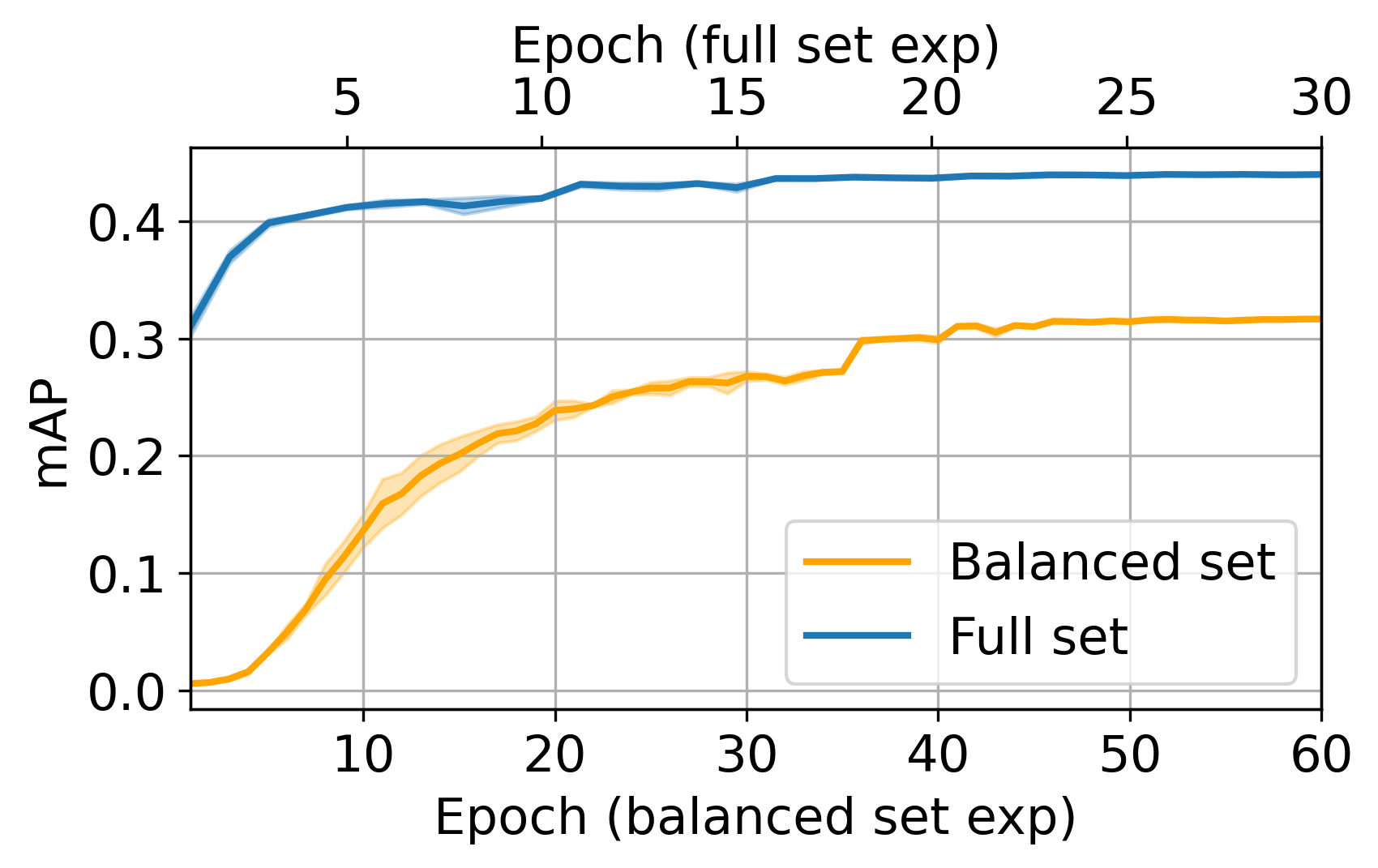}
  \caption{The learning curve of our experiments. Each experiment is run three times, and the stand deviation is shown in the shade.}
  \label{fig:stable}
\end{figure}

We show the learning curve of our best single EfficientNet B2 with 4-headed attention model (without weight averaging) in Figure~\ref{fig:stable}. For both the balanced set and full set experiment, we repeat the training process three times with different random seeds and show the standard deviation in the plot. As we can see, the training converges, and the performance of the model barely varies with the random seed, i.e., the three runs achieve almost the same result.

\section{Conclusion}
\label{sec:discuss}

In this paper, we describe several techniques that improve the performance of a CNN-based neural model for audio tagging.  First, we show an ImageNet-pretrained CNN can noticeably improve performance. While it is straightforward to implement for CNN-based models it has seldom been used in audio tagging research.  Second, due to an imbalance in sound class samples in Audioset, we describe several data balancing and augmentation strategies that alleviate the data imbalance issue and help improve performance. We argue that balanced sampling and data augmentation should be a standard component for AudioSet modeling. Third, by observing variation in class-specific performance, we identified a missing label issue with Audioset and proposed a label enhancement method that shows improvement on the balanced training set. The enhanced label set can be used in the same way as the original label set in future research. We were not able to observe a performance improvement by enhancing the full set labels, possibly due to similar missing labels in the evaluation set. Due to its impact on performance, we believe addressing the noisy label issue is an important research topic for audio tagging. Finally, we describe weight averaging and ensemble strategies that are both simple and effective for audio tagging.

By combining all these training techniques, we are able to improve the performance of a normal EfficientNet model by 130.6\% and 28.2\% without modifying the model architecture for the balanced and full AudioSet experiment, respectively. This magnitude of improvement is larger than was achieved by many previous model architecture or attention module development efforts, indicating that appropriate training techniques are equally important. As a consequence, by training an EfficientNet with these techniques, we obtain a single model (with 13.6M parameters) and an ensemble model that achieve mean average precision (mAP) scores of 0.444 and 0.474 on AudioSet, respectively, outperforming the previous best system of 0.439 with 81M parameters~\cite{kong2020panns}. Our best model trained with only the balanced AudioSet ($\sim1\%$ of the full set) outperforms our baseline and many previous models trained with the full set. We show the AUC and d-prime of our models and compare them with previous efforts in Table~\ref{tab:final}. The proposed model outperform previous models for all evaluation metrics.

\begin{table}[t]
\centering
\caption{Comparison with Previous Methods (Upper: Balanced AudioSet Experiments, Lower: Full AudioSet Experiments).}
\label{tab:final}
\begin{tabular}{@{}lcccc@{}}
\toprule
                        & \#Params & mAP                                                      & AUC                                                      & $d^\prime$                                                \\ \midrule

Wu-minimal~\cite{wu2018reducing}, 2018              & 2.6M           & -                                                        & 0.916                                                    & 1.950                                                  \\
Kumar~\cite{kumar2018knowledge}, 2018                   &      -         & 0.213                                                    & 0.927                                                    & 2.056                                                  \\
Wu-best~\cite{wu2018reducing}, 2018                 & 56M            & -                                                        & 0.927                                                    & 2.056                                                  \\
Kong~\cite{kong2019weakly}, 2019                    & -             & 0.274                                                    & 0.949                                                    & 2.316                                                  \\
PANNs~\cite{kong2020panns}, 2020                   & 81M            & 0.278                                                    & 0.905                                                    & 1.853                                                  \\ \midrule
Our Baseline     & 13.6M            & 0.1570                                                   & 0.9108                                                   & 1.903                                                  \\
Proposed Single Model        & 13.6M            & \begin{tabular}[c]{@{}c@{}}0.3192\\ $\pm$0.0015\end{tabular} & \begin{tabular}[c]{@{}c@{}}0.9534\\ $\pm$0.0005\end{tabular} & \begin{tabular}[c]{@{}c@{}}2.374\\ $\pm$0.007\end{tabular} \\
Proposed 68M Model         & 13.6M$\times$5          & 0.3527                                                   & 0.9602                                                   & 2.479                                                  \\
Proposed Full Model   & 13.6M$\times$20         & {\bf 0.3620}                                                   & {\bf 0.9638}                                                   & {\bf 2.541}                                                  \\ \midrule \midrule
AudioSet Baseline~\cite{gemmeke2017audio} & -             & 0.314                                                    & 0.959                                                    & 2.452                                                  \\
Kong~\cite{kong2018audio}, 2018              & -             & 0.327                                                    & 0.965                                                    & 2.558                                                  \\
Yu~\cite{yu2018multi}, 2018                      &    -           & 0.360                                                    & 0.970                                                    & 2.660                                                  \\
TALNet~\cite{wang2019comparison}, 2019                     &      -         & 0.362                                                    & 0.965                                                    & 2.554                                                  \\
Kong~\cite{kong2019weakly}, 2019                    &     -          & 0.369                                                    & 0.969                                                    & 2.639                                                  \\
DeepRes~\cite{ford2019deep}, 2019                 & 26M            & 0.392                                                    & 0.971                                                    & 2.682                                                  \\
PANNs~\cite{kong2020panns}, 2020                   & 81M            & 0.439                                                    & 0.973                                                    & 2.725                                                  \\ \midrule
Our Baseline         & 13.6M            & 0.3723                                                   & 0.9706                                                   & 2.672                                                  \\
Proposed Single Model       & 13.6M            & \begin{tabular}[c]{@{}c@{}}0.4435\\ $\pm$0.0008\end{tabular} & \begin{tabular}[c]{@{}c@{}}0.9753\\ $\pm$0.0003\end{tabular} & \begin{tabular}[c]{@{}c@{}}2.778\\ $\pm$0.007\end{tabular} \\
Proposed 68M Model        & 13.6M$\times$5          & 0.4690                                                   & 0.9789                                                   & 2.872                                                  \\
Proposed Full Model  & 13.6M$\times$10         & {\bf 0.4744}                                                   & {\bf 0.9810}                                                   & {\bf 2.936}        \\ \bottomrule
\end{tabular}
\end{table}

The work in this paper can serve as a recipe for AudioSet training. Most of the proposed methods are model agnostic and can be combined together with various model architectures and attention modules. As we showed in the paper, the same model can perform much better when it is trained with appropriate techniques. We hope this work can facilitate future audio tagging research by documenting a set of strong and useful training techniques.

\section{Acknowledgment}
This work was supported in part by Signify.


\ifCLASSOPTIONcaptionsoff
  \newpage
\fi

\bibliographystyle{IEEEtran}
\bibliography{ref}

\end{document}